\documentclass[a4paper,fleqn,usenatbib]{mnras}
\usepackage{mathptmx}
\usepackage{pdflscape}

\usepackage[T1]{fontenc}
\usepackage{ae,aecompl}

\usepackage{graphicx}	
\usepackage{amsmath}	
\usepackage{amssymb}	

\title[]{A SITELLE view of M31's central region -- I: Calibrations and
  radial velocity catalogue of nearly 800 emission-line point-like
  sources}

\author[Thomas B. Martin et al.]{
Thomas B. Martin$^{1,2}$\thanks{E-mail: thomas.martin.1@ulaval.ca},
Laurent Drissen$^{1,2,3}$,
Anne-Laure Melchior$^{4}$
\\
$^{1}$D\'epartement de physique, de g\'enie physique et d'optique, Universit{\'e} Laval, 2325, rue de l'universit{\'e}, Qu{\'e}bec (Qu{\'e}bec), G1V 0A6, Canada\\
$^{2}$Centre de recherche en astrophysique du Québec\\
$^{3}$Department of Physics and Astronomy, University of Hawai'i at Hilo, Hilo, HI 96720, USA\\
$^{4}$LERMA, Sorbonne Universités, UPMC Univ. Paris 6, Observatoire de Paris, PSL Research University, CNRS, 75000, Paris, France
}

\date{Accepted XXX. Received YYY; in original form ZZZ}

\pubyear{2016}

\DeclareRobustCommand{\Ha}{\textup{H}$\alpha$}
\DeclareRobustCommand{\HII}{\textup{H\,\textsc{\lowercase{II}}}}
\DeclareRobustCommand{\NII}{\textup{[N\,\textsc{\lowercase{II}}]}}
\DeclareRobustCommand{\SII}{\textup{[S\,\textsc{\lowercase{II}}]}}
\DeclareRobustCommand{\OIII}{\textup{[O\,\textsc{\lowercase{III}}]}}

\DeclareRobustCommand{\kms}{km\,s$^{-1}$}
\begin{document}
\label{firstpage}
\pagerange{\pageref{firstpage}--\pageref{lastpage}}
\maketitle

\begin{abstract}
  We present a detailed description of the wavelength, astrometric and
  photometric calibration plan for SITELLE, the imaging Fourier
  transform spectrometer attached to the Canada-France-Hawaii
  telescope, based on observations of a red (647 - 685 nm) data cube
  of the central region (11$' \times 11'$) of M\,31. The first
  application, presented in this paper, is a radial-velocity catalogue
  (with uncertainties of $\sim 2 - 6$ km/s) of nearly 800
  emission-line point-like sources, including $\sim$ 450 new
  discoveries.  Most of the sources are likely planetary nebulae,
  although we also detect five novae (having erupted in the first 8
  months of 2016) and one new supernova remnant candidate.
\end{abstract}




\section{Introduction}

Aiming at characterizing the nearest liner, at the core of M\,31, by
studying line ratios and kinematics of the diffuse gas surrounding it
(Melchior et al., in preparation), we have obtained a medium
resolution (R $\sim$ 5000) data cube in the \Ha{}--\NII{}--\SII{}
range (649--684\,nm) of a large section around the
galaxy nucleus with the imaging Fourier transform spectrometer (iFTS)
SITELLE. 

This instrument, described in \citet{Baril2016, Drissen2010,
  Grandmont2012}; Drissen et al. (in prep.), consists of a Michelson
interferometer inserted into the collimated beam of an astronomical
camera system, equipped with two fast-readout, low noise, deep
depletion e2v CCD detectors ($2048 \times 2048$, $0.32''$ pixels)
providing a total field of view of $11' \times 11'$. The spectral
range of a given data cube is selected by inserting the desired
filter, chosen from a list covering the entire visible range, from 350
nm to 850 nm, in front of the collimator. SITELLE's spectral
resolution is very flexible, ranging from R = 1 (panchromatic) to R =
20 000. Details on data acquisition, and the process of transforming
the two initial interferometric cubes into a single, calibrated
spectral cube, are presented below.

The detection of several hundred
\Ha{}-emitting point sources in our M31 cube, some of which being new discoveries, has
motivated the creation of a catalogue and, for this purpose, a
significant improvement of the calibration methods used for SITELLE's
first data release (\citealt{Martin2016b}, Martin et al. in
preparation).

After a brief description of the instrument and the observations, Section 2 of this paper
describes in details the wavelength, astrometric, and flux calibration
of SITELLE data. Section 3 presents the method used to detect the
\Ha{}-emitting point sources before introducing the catalogue and
a comparison with previous work based on narrow-band imagery and
multi-object dispersive spectroscopy.

\section{Observations and Data Calibration}
\subsection{Observations}
The data cube was obtained on August 24, 2016, with SITELLE attached
to the Canada-France-Hawaii telescope, and the SN3 filter
(647--685\,nm), designed to detect the \Ha{} line as well as the
\NII{} 6548, 6584 and \SII{} 6717,6731 doublets in Milky Way {\HII}
regions and nearby galaxies up to z=0.017. Parameters of the
observations are listed in Table~\ref{tab:obsparams}. The duration of
the cube was 4.1 hours, including the 3.8-s overhead per step (CCD
readout and mirror displacement and stabilization) giving a total
on-source exposure time of 3.2 hours; sky was photometric and the
median seeing, $0.9\,\arcsec$, was well sampled by the $0.32\,\arcsec$
CCDs attached to both output ports of the
interferometer. The target resolution was 5000 on the interferometer
axis giving a resolution of 4900 at the center of the field of view
(at an incident angle of 15.4$^{\circ}$). As we have not measured
emission lines with a broadening significantly smaller than
$\sim$20\,\kms, we consider that this may be due to a loss of
resolution of 1.7\,\%, i.e. an effective resolution of 4800. The data
has been reduced with ORBS \citep{Martin2012, Martin2015,
  Martin2015-thesis}. Wavelength calibration has been performed with a
method developed for ORCS \citep{Martin2015} and described in details
in section~\ref{sec:wave_calib_method}.

Although SITELLE was designed to reach higher spectral resolutions and
that we have demonstrated that it can also reach this resolution
during Commissioning and Science verification phases (Drissen et al.,
in prep.), this is only the second time that such a long scan has been
obtained with the instrument. It was therefore an ideal opportunity to
improve SITELLE's calibration routine and compare our data with
independent studies, all the more so that a large number of
\Ha{}-emitting point sources, most of which had previously published
coordinates and radial velocities, were easily detected in the cube.
 
\begin{table}
  \caption{Observation parameters of M\,31. R is the effective
    resolution measured at the center of the data cube at the
    wavelength of \Ha.}
  \label{tab:obsparams}
  \begin{tabular}{lcccccc}
    \hline
    Filter & Exposure& Folding&Step& Step& ZPD & R \\
           &time/step & order& size&nb.&&\\
    \hline
    SN3 & 13.7\,s &8& 2943\,nm&840&168&4800\\
    \hline
  \end{tabular}
\end{table}

\subsection{Wavelength calibration}

\label{sec:wavelength_calibration}

SITELLE's observation method is described in detail in
\citealt{Drissen2010} and \citealt{Martin2016}. An interferometric
cube is obtained by moving one of the interferometer's mirrors, while
keeping the other one at a fixed position. The optical path difference
(OPD) between the two interfering beams thus changes and modulates the
intensity measured at the output port of the interferometer.

\begin{figure}
  \includegraphics[width=\columnwidth]{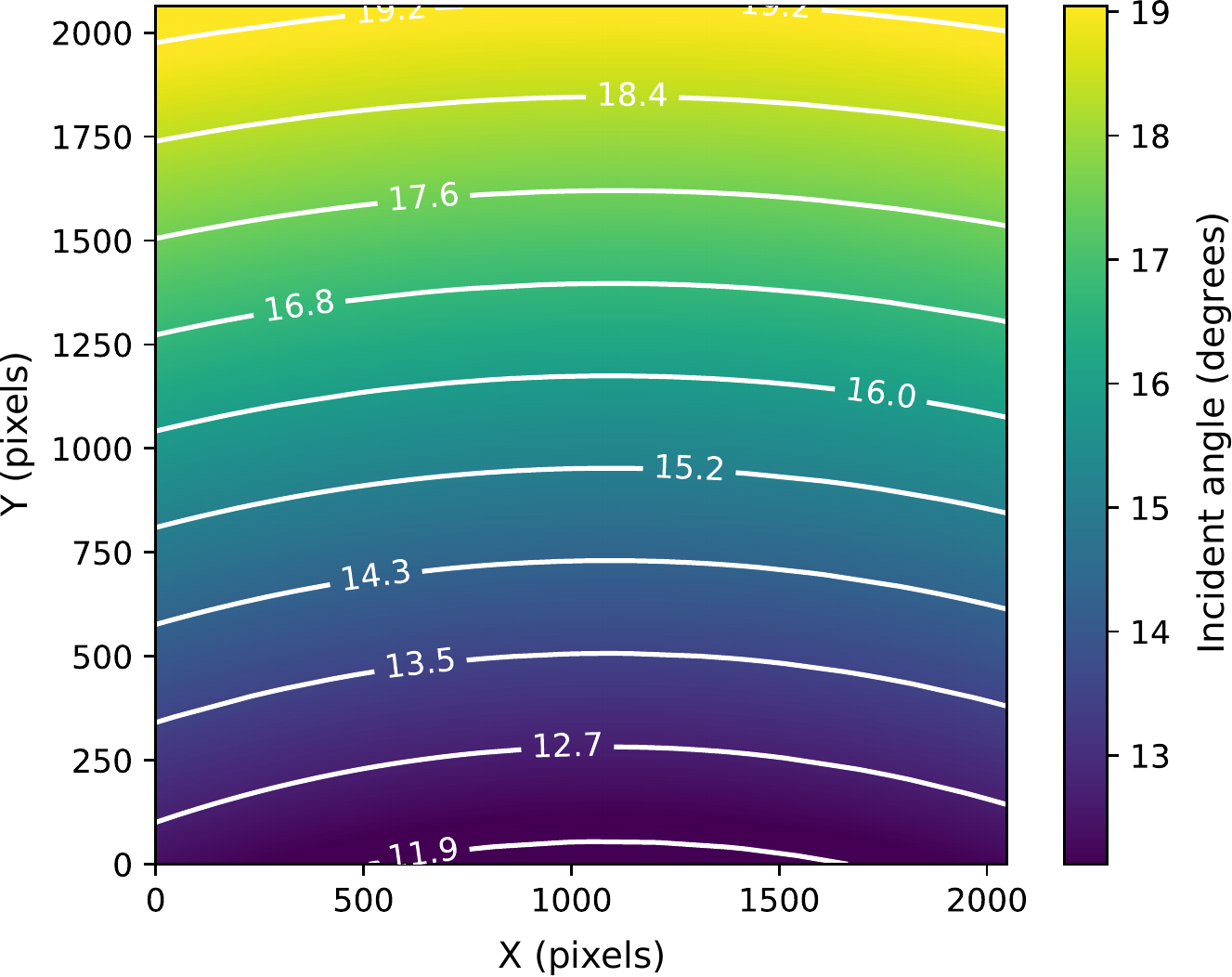}
  \caption{Incident angle ($\theta$) in the interferometer at each
    pixel. This angle was determined from the central wavelength of a
    HeNe (543 nm) laser as measured in SITELLE with a high resolution
    data cube; it varies between 11.8$^{\circ}$ and 19.6$^{\circ}$. The
    ``observed'' wavelength of the laser thus varies between ~555 nm
    (543 nm / cos[11.8$^{\circ}$]) and ~576 nm (543 nm / cos[19.6$^{\circ}$]). }
    \label{fig:laser_map_orig}
\end{figure}

SITELLE is based on an off-axis interferometer configuration in order
to collect the entire flux from the source from the combination of two
complementary output ports instead of one in the classical Michelson
configuration. The center of the field of view is thus 15.5$^o$ away
from the interferometer axis. On an interferometric image, each pixel
measures the flux at a given incident angle $\theta$, with respect to
the interferometer axis where $\theta=0$. Because of the off-axis
configuration, $\theta$ varies between 11.8$^o$ and 19.6$^o$
(Fig.~\ref{fig:laser_map_orig}).

During a scan, the moving mirror is moved away from its original
position, the Zero Path Difference (ZPD), where the OPD is null, by
steps of equal size. At each step, an interferometric image is
recorded on each output port with two 2k$\times$2k deep-depletion e2v
CCD detectors (named Cam1 and Cam2). The image recorded on Cam2 is
aligned with the image on Cam1 and both images are combined during the
reduction process.

When the interferogram recorded by one pixel is Fourier transformed to
a spectrum, the knowledge of the exact incident angle is enough to
produce the absolute wavelength calibration (\citealt{Martin2016b},
Martin et al. in preparation).  Indeed, as long as the interferogram
is equally sampled, there is no uncertainty on the relative wavelength
calibration. This assertion holds if the distribution of the sampling
error has a mean of zero and its standard deviation is much smaller
than the step size (Martin et al. in preparation). With SITELLE,
wavelength calibration relies on a high resolution cube of a laser
source obtained with the telescope pointing at zenith. The laser
source is a green He-Ne laser at 543.5\,nm. The exact wavelength of
the laser is not very well known and is subject to drift with
time. The absolute calibration of the whole cube is thus subject to a
systematic bias but the relative, pixel-to-pixel, calibration is not
affected. However, the relative calibration will show some distortions
from a perfect model due to aberrations and deformations in the
optical structure.

Even though a lot of efforts have been put on its robustness to
temperature variations and changes of the direction of the gravity
vector, the structure is still not ideal. By obtaining laser cubes in
different directions with a zenithal angle of 47\,$^{\circ}$
(corresponding to an airmass of 1.5), we have assessed during
SITELLE's commissioning, that the deformation of the optical structure
could lead to relative velocity errors of up to 15\,\kms{}, depending
on the location of the target in the sky.  The difference between the
mean direction of the observed target and the zenith calibration laser
cube being always smaller than 47\,$^{\circ}$, the errors on the
velocity measurements are expected to be less than 15\,\kms{}, even
with this basic calibration procedure. It is possible to refine this
calibration by measuring the velocity of the night sky emission lines,
when present in the filter bandpass. In particular, the SN3 filter
(647 -- 685 nm), which has so far been used to obtain the highest
spectral resolution cubes with SITELLE, includes numerous Meinel OH
bands. We have implemented a method in ORCS, SITELLE's dedicated data
analysis software, \citep{Martin2015} to compute a map of the sky
lines velocity in the field of view which has already been used to
improve the wavelength calibration of other data sets
\citep{Martin2016, Shara2016}. We have refined this method to
calibrate the data obtained on M\,31. Details are provided in the next
section.

\subsubsection{Method}
\label{sec:wave_calib_method}

Up to now, measuring sky lines velocity has proven to be the best way
to calibrate a cube, both relatively and absolutely, regardless of the
initial wavelength calibration. But this method has always been used
on spectral cubes where the sky lines are detected with a decent SNR
over the whole field of view. In those cases, sky spectra are
extracted everywhere in the field and, eventually, a bi-spline model
is fitted to interpolate the velocity at each pixel of the cube (see
section~\ref{sec:fit_of_the_sky_velocity}).

However, in the case of M\,31, the high level of continuum near the
center of the object enhanced by the multiplex disadvantage of the FTS
technique \footnote{The multiplex disadvantage comes from the fact
  that all the photons of the interferogram ($\sim N$ photons per
  exposure $\times$ $M$ exposures) are used to compute each spectral
  sample leading to a photon noise on each of the $M$ spectral
  channels equal to the photon noise coming from $N \times M$ photons
  (i.e.  $\sqrt{N\times M}$). If a dispersive technique was used to
  acquire the exact same spectrum with the same exposure time on a
  continuum source (e.g. a star emitting $N$ photons per channel) the
  photon noise per channel would be $\sim\sqrt{N}$ i.e. $\sqrt{M}$
  time better than the FTS technique. Note that this disadvantage is
  largely reduced if the source is monochromatic.}
\citep{Drissen2012,Drissen2014,Maillard2013}, combined with the short
exposure time per step, strongly reduces the SNR and prevents any
precise ($\leq 2$ km/s) measurement of the sky lines centroid in the
center of the field of view (FOV). Nevertheless, on the border of the
FOV, the level of continuum is low enough to provide accurate
measurements of the sky lines velocity (see
Figure~\ref{fig:sky_map_all}). Since the wavelength calibration is
strongly tied to the optical structure of the interferometer, we can
model the wavelength calibration map from a small set of instrumental
parameters and fit the known velocity points (i.e. on the border of
the FOV) to estimate the velocity where no reliable measure of the sky
lines can be made (i.e. at the center FOV). It is then possible to
compute the correct wavelength calibration everywhere in the FOV with
precision.

This improved calibration method involves:
\begin{enumerate}
\item Computation of a set of initial parameters from the
  calibration laser map measured at zenith;
\item Measurement of the velocity of the sky lines, which
  translates directly into a wavelength calibration error since this
  velocity must be 0;
\item Modeling of a new calibration laser map with a new set of
  parameters which fit the measured velocity of the sky lines.
\end{enumerate}

We will now describe the model of the calibration laser map we have
developed to relate the deformation of the optical structure to the
value of the incident angle at each pixel of the FOV. Note
that the calibration laser map is taken at zenith. The fit of this
calibration laser map with the model permits to define an initial set
of parameters which can eventually be changed to model what would be
the real calibration laser map in the direction of the target (here
M\,31).

\subsubsection{Calibration laser map modelling}
\label{sec:calibration_laser_map_modelling}

The best way to fit a calibration laser map is to construct a model
$M$ defined as a function of instrumental parameters $p$, which
permits to calculate $\sigma$, the measured laser wavenumber related
to the wavelength $\lambda$ via $\sigma=1 / \lambda$, at any given
pixel position ($x$, $y$).
\begin{equation}
  \label{eq:model_base}
  \sigma (x,y) = M(x, y, p)\;,
\end{equation}

Recalling that the angle between the interferometer axis and the
detector pixel is $\theta$, we can start by providing the relation
between the measured velocity error $\epsilon_v$ and the error
$\epsilon_\theta$ made on the calculated value of the incident angle
$\theta$. Indeed, following the equation~\ref{eq:sigma_cos_theta}, any
error on $\theta$ translates into an error on the measured wavenumber,
$\sigma$, with respect to the real wavenumber $\sigma_0$,
\begin{equation}
\label{eq:sigma_theta}
\sigma = \frac{\sigma_0}{\cos(\theta + \epsilon_\theta)}\;.
\end{equation}
This error will in turn result in an error on the measured velocity,
\begin{equation}
  \epsilon_v = c \left(\frac{1}{\cos(\theta + \epsilon_\theta)} - \frac{1}{\cos(\theta)}\right)\;,
\end{equation}
where $c$ is the speed of light.

\begin{figure*}
  \includegraphics[width=\linewidth]{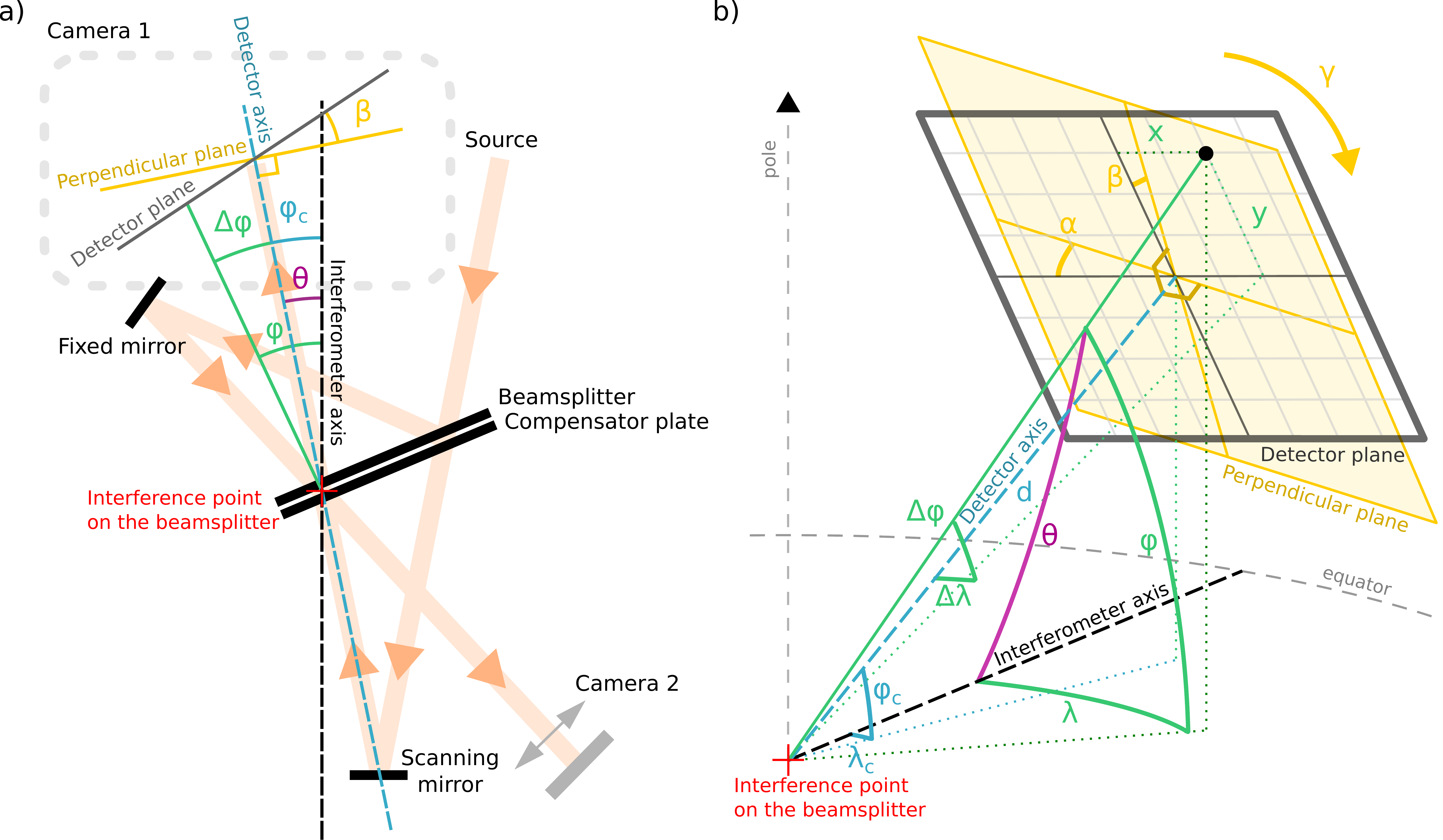}
  \caption{Idealized model of the interferometer
    structure. \textit{a)} 2D projection of the whole interferometer
    structure. The off-axis configuration, with a beamsplitter tilted
    with respect to the classical 45$^{\circ}$ configuration, appears
    clearly. Some of the indications shown in the complete 3D
    figure (b) are reported. The nominal incident angle $\theta$ at
    the center of the detector is 15.5$^{\circ}$. By projection the
    angles $\theta$ and $\varphi$ seem to be the same but they are
    not. Note that the optics of the cameras, in front of the detector
    plane, are not taken into account in the model. \textit{b)} 3D
    representation of the model of the interferometer. The equator
    plane (or reference plane) and the pole direction of the spherical
    coordinates system are indicated in grey. The origin is the point
    on the beamsplitter where the detector axis crosses the
    semi-transparent coating.}
    \label{fig:coords}
\end{figure*}

If the optics in front of the detector are not taken into account,
i.e. the computed distances are not corrected for the optical
magnification, an idealized structural model can be deduced from
simple geometrical considerations (the optics and the detector are
also considered distortion free). The detector is considered as a
perfect plane perpendicular to the direction of the incoming beam at a
distance $d$ from the beamsplitter. Projected into spherical
coordinates, as shown in Fig.~\ref{fig:coords}, this angle is defined
by its longitude $\lambda$ and its latitude $\varphi$. The angles that
define the direction from the origin to the center of the detector
(the detector axis) are $\lambda_{\text{c}}$ and $\varphi_{\text{c}}$.

To calculate the incident angle of any pixel on the detector, one can
first determine the angles $\Delta\lambda$ and $\Delta\varphi$ of this
pixel with respect to the detector axis (the direction which points
from the origin to the center of the detector). Since the physical
size of the pixel $p$ is known (15 $\mu$m), the physical position of a
pixel on the detector ($x$, $y$) can be simply related to its index
($i_x$, $i_y$)
\begin{gather}
  \label{eq:xy}
  x = p\,i_x\\
  y = p\,i_y
\end{gather}
The rotation of the detector by an angle $\gamma$ can be taken into
account with a rotation matrix
\begin{gather}
  \label{eq:xy_rot}
  x = p\,i_x \cos(\gamma) + p\,i_y \sin(\gamma) \\
  y = p\,i_y\cos(\gamma) - p\,i_x \sin(\gamma)
\end{gather}
The plane of the detector can also be tilted with respect to the plane
perpendicular to the detector axis (tip-tilt angles $\alpha$ and
$\beta$). The angle $\Delta\lambda$ and $\Delta\varphi$ are computed
with the classical tangent law
\begin{gather}
  \label{eq:x2theta}
  \Delta\lambda = \arctan\left(\frac{x - d}{x + d}\frac{1}{\tan(0.5(\alpha + \pi/2))}\right) - \frac{\alpha}{2} + \frac{\pi}{4}\\
  \Delta\varphi = \arctan\left(\frac{y - d}{y + d}\frac{1}{\tan(0.5(\beta + \pi/2))}\right) - \frac{\beta}{2} + \frac{\pi}{4}
\end{gather}
The direction of the pixel with respect to the interferometer axis is thus
\begin{gather}
  \lambda = \lambda_c + \Delta\lambda \\
  \varphi = \varphi_c + \Delta\varphi
\end{gather}
and the incident angle $\theta$ can finally be derived with Vincenty's
formula \citep{Vincenty1975},
\begin{equation}
  \label{eq:theta}
  \theta = \arctan\left(\frac{\sqrt{(\cos(\varphi)\sin(\lambda))^2 + \sin(\varphi)^2}}{\cos(\varphi)\cos(\lambda)}\right)\;.
\end{equation}

Finally, from equation~\ref{eq:opd_cos_theta}, we can compute the
measured wavenumber of the laser source of nominal wavenumber
$\sigma_0$ at a given incident angle $\theta$ (we recall the equation
here for clarity)
\begin{equation}
  \sigma_{\theta} = \sigma_0\cos(\theta)\;.
\end{equation}

This model provides a good description of the measurement of the
incident angle that can be done via the observation of a calibrated
laser source. If we look at the typical values of the parameters
returned by a fit of an arbitrary calibration laser map we find a
distance $d$ of $\sim$23.5\,cm (depending on the chosen wavelength of
the calibration laser), which, magnified by 3.3 (the magnification of
the camera optics), gives a real distance of $\sim$76\,cm: a few
centimeters larger than the real distance from the dielectric coating
to the detector surface. Tip-tilt angles are also generally small
which appears mechanically correct.

\begin{figure}
  \includegraphics[width=\columnwidth]{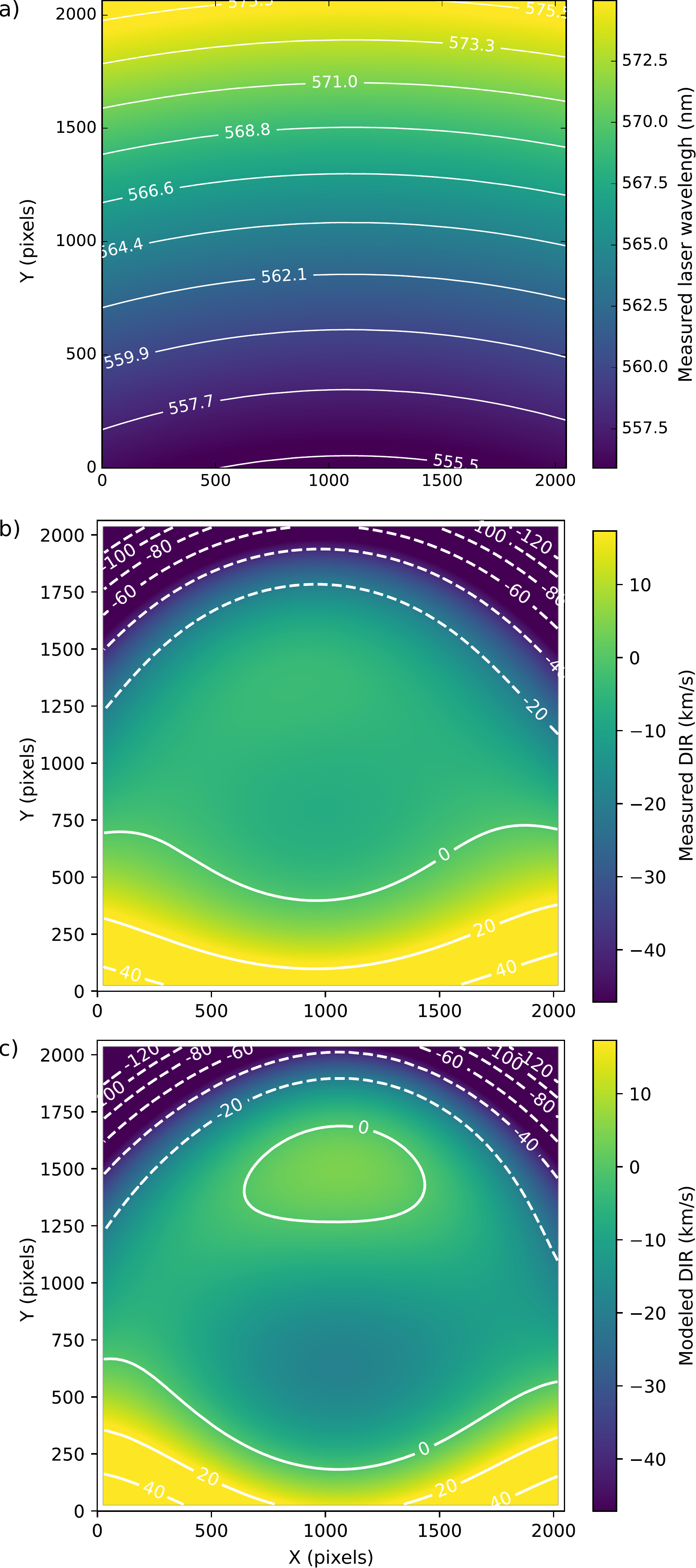}
  \caption{\textit{a)} Calibration laser map. The real wavelength of
    the calibration laser is 543.5\,nm. The measured wavelength
    depends on the incident angle $\theta$ of the incoming light with
    respect to the interferometer axis (see Fig. 1). \textit{b)}
    Distortion Induced Residual once the structural model is
    subtracted. The wavelength error is given in terms of velocity
    error at \Ha{}.  \textit{c)} Modeled DIR computed with the
    astrometric distortion model. Contours are solid when the value is
    positive and dashed when negative.  }
    \label{fig:laser_map_all}
\end{figure}

However,this model does not take into account the distortions produced
by the optics which slightly changes the position where the light at a
given incident angle is measured on the detector (see
Fig.~\ref{fig:laser_map_all}).

If we assume that our model is correct in first approximation, the residual from
the fit on the measured calibration laser map must come from optical
aberrations, such as distortions and wavefront errors, with some normally
distributed noise, $\mathcal{N}$, that can be reduced by fitting the
residual with an appropriate model (Zernike polynomials in this case),
i.e., from equation~\ref{eq:model_base}:
\begin{equation}
  \label{eq:model_true}
    \sigma_{\text{observed}}(x, y) = M(x, y, p) + DIR(x, y) + \mathcal{N}\;,
\end{equation}
DIR$(x,y)$ being the Distortion Induced Residual, as the optical
distortions are a major contributor to this residual. Indeed, we can
use the distortion model obtained from the astrometric calibration
(see section~\ref{sec:astrometric_calibration}) and try to model what
would be the observed DIR (see Fig.~\ref{fig:laser_map_all}). We can
see that, even if small features are not perfectly reproduced, most of
the measured DIR can be explained by optical distortions, especially
in the corners. Note also that we are using a distortion model
calculated in the red part of the spectrum (the SN3 filter is centered
at 666 nm) while the calibration laser map is observed at 543.5\,nm
and that distortions are likely to be chromatic. A more careful study
of the relation between the residual and the astrometric distortion
pattern still needs to be done but is beyond the scope of this paper.

The idea that a complete model of the calibration map is the sum of a
geometrical model of the interferometer that changes with the
direction of the gravity vector plus a constant DIR (see
equation~\ref{eq:model_true}) is confirmed by the comparison of the
DIRs computed from the fit of 5 calibrations laser maps obtained at
different telescope pointings. During the commissioning, 4 calibration
laser maps have been obtained at an angle of 47\,$^{\circ}$ in 4
different directions (north, south, east, west). They have been
compared with a calibration laser map obtained at zenith. We have
found that the difference between the DIRs computed with the fit of
each of these 5 calibration laser maps was always smaller than
1\,\kms{}. However, the difference between the geometrical models
$M(x,y,p)$ can reach 20\,\kms{}. Note that the final residual made on
each fit (which, from equation~\ref{eq:model_true}, must be noise
only) is a perfect Gaussian distribution with a standard deviation of
0.5\,\kms{}. We conclude that, as long as the optics are not changed,
the DIR remains stable and the real calibration map of a target
observed in a direction different from the zenith only depends on the
instrumental parameters $p$ of the geometrical model $M(x,y,p)$. Note
that a standard deviation of 0.5\,\kms{} on the fit of the calibration
laser map does not mean that our precision on the wavelength
calibration will be the same. This result better gives a lower limit
on the calibration precision. The wavelength calibration of the actual
data is discussed in the next section.

\subsubsection{Fit of the sky velocity}
\label{sec:fit_of_the_sky_velocity}

\begin{figure}
  \includegraphics[width=\columnwidth]{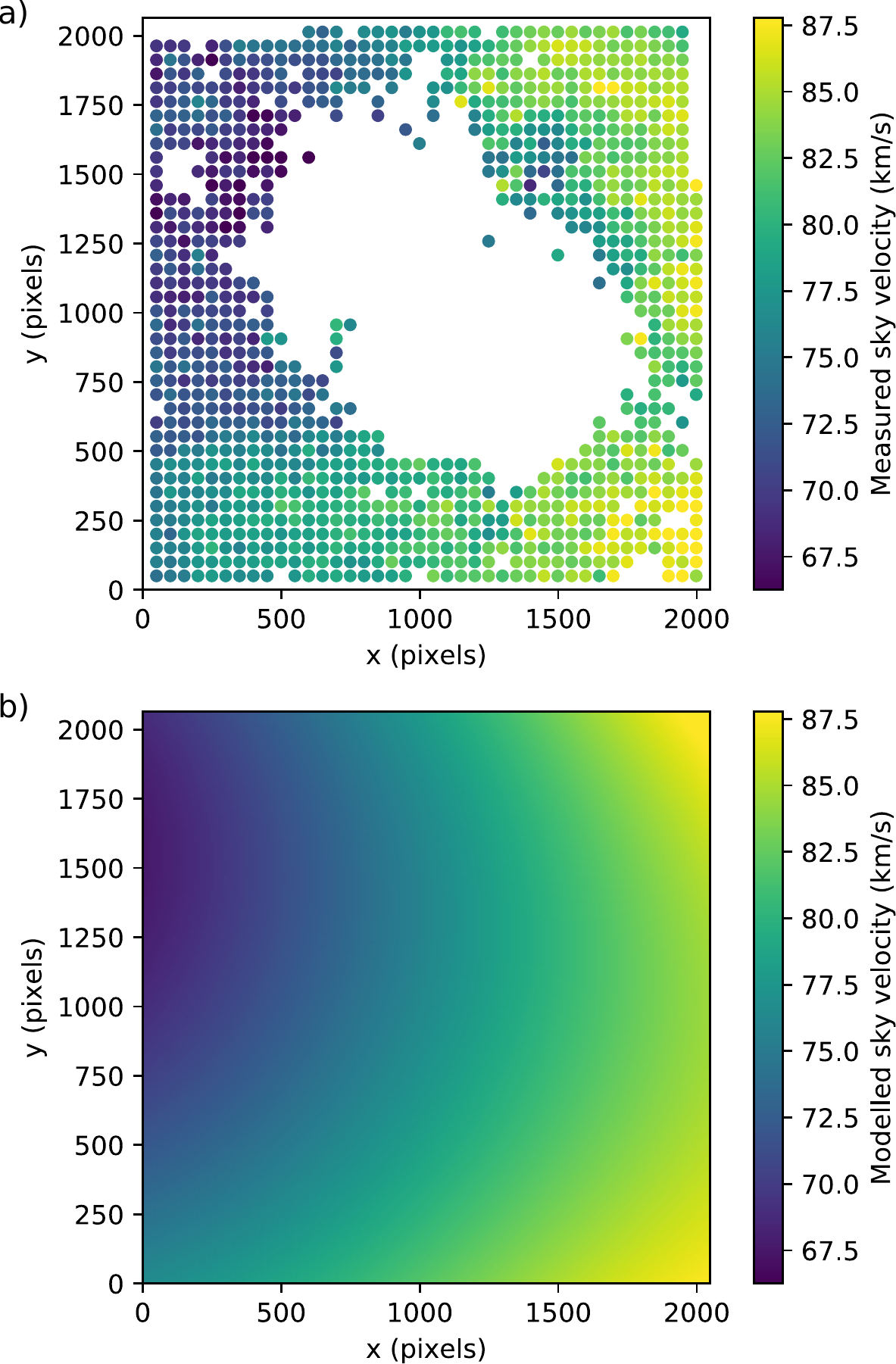}
  \caption{\textit{a)} Measured velocity of the sky lines in the SN3
    filter on integrated spectra of 30$\times$30 pixels in SITELLE's
    field of view. The points where no sky lines could be measured and
    the points having an uncertainty greater than 1.7\,\kms are not
    displayed (and were not used for the fit). The white area
    (depopulated from reliable measurements) corresponds to the
    central region, where the continuum flux is very high. The main
    goal of the fitted model shown below is to give an estimate of the
    sky velocity in this area (b) \textit{b)} Velocity model of the
    sky lines computed from the calibration laser map model and fitted
    over the velocity points shown in (a). Fitted parameters of the
    model are reported in Table~\ref{tab:wave_calib_params} and
    residuals of the fit are shown in Figure~\ref{fig:residual_all}.}
    \label{fig:sky_map_all}
\end{figure}

We have measured the velocity of the sky lines over the whole field of
view by extracting 1600 spectra at each point of a 40$\times$40
grid. Each sky spectrum was integrated over a 30$\times$30 pixels box
to improve its signal-to-noise ratio. But, as explained in
section~\ref{sec:wave_calib_method} the high level of continuum near
the center of the galaxy prevents any precise measure over a large
part around the center of the field (see
Fig.~\ref{fig:sky_map_all}). We have thus used the model described in
the previous section to compute the real instrumental parameters
during the observation of M\,31 by fitting our model to the measured
velocity of the sky lines (Table~\ref{tab:wave_calib_params}). Note
that the rotation angle has been fixed at 0 because there is a strong
degeneracy between the rotation angle and $\varphi_c$ which leads to
unphysical values of the structural parameters without giving an
appreciable difference in the precision of the fit.

\begin{figure}
  \includegraphics[width=\columnwidth]{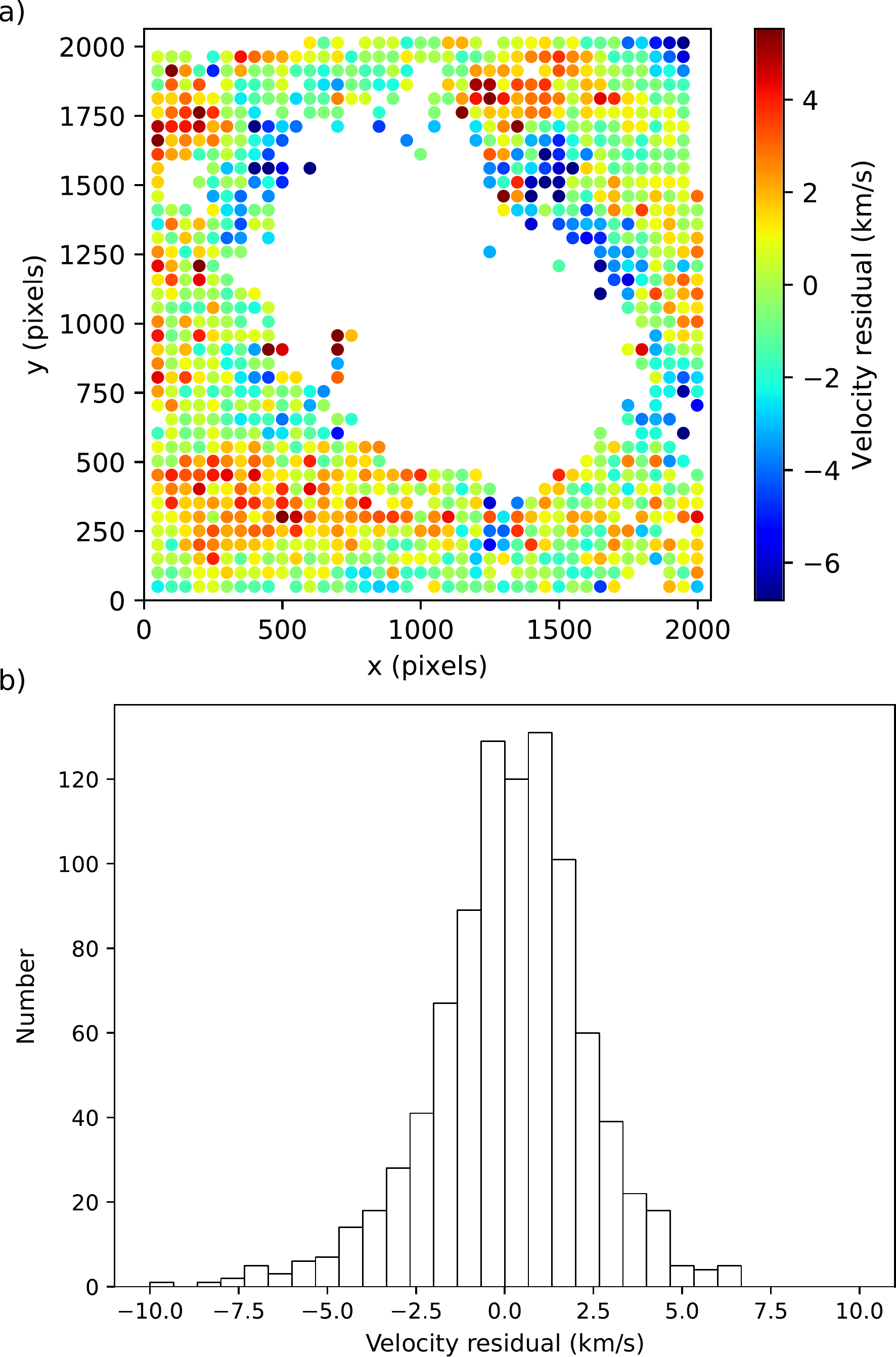}
  \caption{\textit{a) }Map of the residual error on the velocity
    calibration after the computed model is removed. \textit{b)
    }Histogram of the residual error on the velocity calibration
    corresponding to the map presented above. The median of the
    distribution is 0.25\,\kms{} and its standard deviation is
    2.21\,\kms{}. This systematic uncertainty has been added to all
    velocity measurements.}
    \label{fig:residual_all}
\end{figure}

With this method we have been able to estimate the sky lines velocity
in the central region of the field of view where no sky lines could be
measured and reduce the initial velocity gradient of more than
15\,\kms{} (Fig.~\ref{fig:sky_map_all}) down to a much flatter error
with a standard deviation of 2.21\,\kms{}
(Fig.~\ref{fig:residual_all}). This is considered to be our systematic
uncertainty and has been added to all the estimations of the
uncertainty on the velocity measurements.

Nevertheless we must admit that, the standard deviation is still
higher than the 0.5\,\kms{} obtained with the direct fit of the
calibration laser map (see
section~\ref{sec:calibration_laser_map_modelling}). We see two
possibilities regarding the accuracy of the DIR. First, the stability
of the DIR, which is not recomputed (only the instrumental parameters
can be fitted), is ensured as long as the optics are not
touched. Sadly, no calibration laser cube has been taken during the
night of the observation, which was the first night of the run. The
next night, one of the two cameras (Cam1) was removed to perform some
tests and was put back in place but with a rotation of 0.8\,$^{\circ}$
(this rotation corresponds to the angle $\theta$ in the model
described in section~\ref{sec:calibration_laser_map_modelling}),
making the measurement of the DIR from the calibration laser, obtained
after the changes on Cam1, much less precise. We have tried to
simulate the real calibration laser map (and thus the real DIR) at the
date when M\,31 was observed by rotating Cam1 on the calibration data
obtained a few days after with no real success. That's why, even if
this possibility cannot be discarded, we do not believe that the
difference we see is due to the camera change. The other possibility
comes from the fact that our model has only been tested with green
laser cubes (543.5\,nm) and it is possible that the DIR calculated at
543.5\,nm may not be exactly the same in the SN3 filter (i.e. around
660\,nm). If we consider that a deviation of 1 pixel in the distortion
pattern can account for up to 6\,\kms{}, small variations of the
refractive index of the beamsplitter could explain why the error is
structured and has a higher than expected standard deviation
(Fig.~\ref{fig:residual_all}).

\begin{table}
  \centering
  \caption{Fitted parameters of the calibration laser map model for
    the wavelength calibration of M\,31 based on the measured sky
    lines velocity. The reference camera is Cam1.}
  \label{tab:wave_calib_params}
  \begin{tabular}{l|l}
    Beamplitter-detector distance($d$)& 23.8\,cm\\
    X angle from the optical axis to the center($\lambda_c$)& -0.47\,$^{\circ}$\\
    Y angle from the optical axis to the center($\varphi_c$)& 15.44\,$^{\circ}$\\
    Tip-tilt angle of the detector along X ($\alpha$)& 0.25\,$^{\circ}$\\
    Tip-tilt angle of the detector along Y ($\beta$)& -0.43\,$^{\circ}$\\
    Rotation angle of the detector ($\gamma$)& 0\,$^{\circ}$ (fixed)\\
    Calibration laser wavelength& 543.37\,nm
  \end{tabular}
\end{table}

\subsection{Astrometric calibration}
\label{sec:astrometric_calibration}
The astrometric calibration model can be divided into three levels of
refinement. The first level starts with the determination of 5 general
registration parameters: the target coordinates in the image, its
celestial coordinates and the angle between the image's Y axis and the
North.  This rough model is arguably more precise near the center of
the frame, where the effects of the optical distortions are
minimal. Indeed, the second level includes a distortion model based on
the Simple Image Polynomial (SIP) convention \citep{Hook2008}. All the
parameters estimated for the first and the second levels of astrometric
calibration can be written into any FITS header \citep{Hanisch2001}
and interpreted with most FITS viewers (e.g. SAO Image
DS9\footnote{\url{http://ds9.si.edu/site/Home.html}}). But, even with
a SIP distortion model, the calibration error can be as large as 1.5\,\arcsec
in the corners of the image (Fig.~\ref{fig:astrom_hist_all} and
Fig.~\ref{fig:astrom_im_all}). The third level of refinement relies on
two distortion maps (one for each axis of the image) which are
computed from the residual between the SIP calibration model and the
real position of the stars in the image; an error on the calibration
smaller than 0.3\,\arcsec (1 image pixel) is then reached. While the first
level only requires a few stars to get an acceptable calibration, the
other two require the observation of a densely populated star field to
precisely determine the distortion model.

\subsubsection{First level: general registration parameters}

The coordinates at the center of the field of view are generally known
with an uncertainty smaller than 15\,\arcsec and the rotation angle cannot
vary by more than 5\,$^{\circ}$. Many optical modifications between
each run since SITELLE's commissioning explain this large uncertainty
on the rotation angle from one observation to another; these are
expected to be much less numerous from now on.  Nevertheless, a good and
robust estimate of the general parameters is obtained with a two step
registration process.
\begin{enumerate}
\item A first reduction of the uncertainty is made from the
  correlation of the 2D histogram of the real list of positions of all
  the stars visible in the image and the positions of the stars found
  in a catalogue. In the case of M\,31, we have used the Gaia DR1
  catalogue\citep{GaiaCollaboration2016a, GaiaCollaboration2016}. With
  this method, the two lists do not need to contain exactly the same
  stars. The peak of the correlation image will provide the shift
  between both lists with an uncertainty which depends on the size of
  the 2D histogram bins (generally of the order of 10 pixels,
  i.e. 3\,\arcsec). Note that a brute force search, which consists in
  a systematic search along a range of possible angles, must first be
  used to find the rotation angle that maximizes the amplitude of the
  histogram peak.
\item A brute force method, which consists in a systematic search
  through all the possible sets of parameters -- each parameter being
  explored over a given range of possible values, is used to find the
  set of parameters which maximizes the total flux measured in the
  image at the positions projected from the catalogue. When a maximum
  of stars are present at those positions the total flux is a its
  highest.
\end{enumerate}

\subsubsection{Second level: SIP distortion model}
Once the general registration parameters are known, it is possible to
match the position projected from a catalogue with the stars on the
image, especially near the center where the error is less than a few
pixels. The parameters of a fourth order SIP distortion model are
found with a least-square Levenberg-Marquardt minimization algorithm
\citep{Levenberg1944,Marquardt1963} to reduce as much as possible the
distance between both lists (which now must contain the same
stars). As the number of suitable stars in M\,31's central region is
small, we have used an image of a field in NGC\,6960, obtained the same
night, to compute the SIP parameters.

\subsubsection{Third level: residual distortion maps}
\begin{figure}
  \includegraphics[width=\columnwidth]{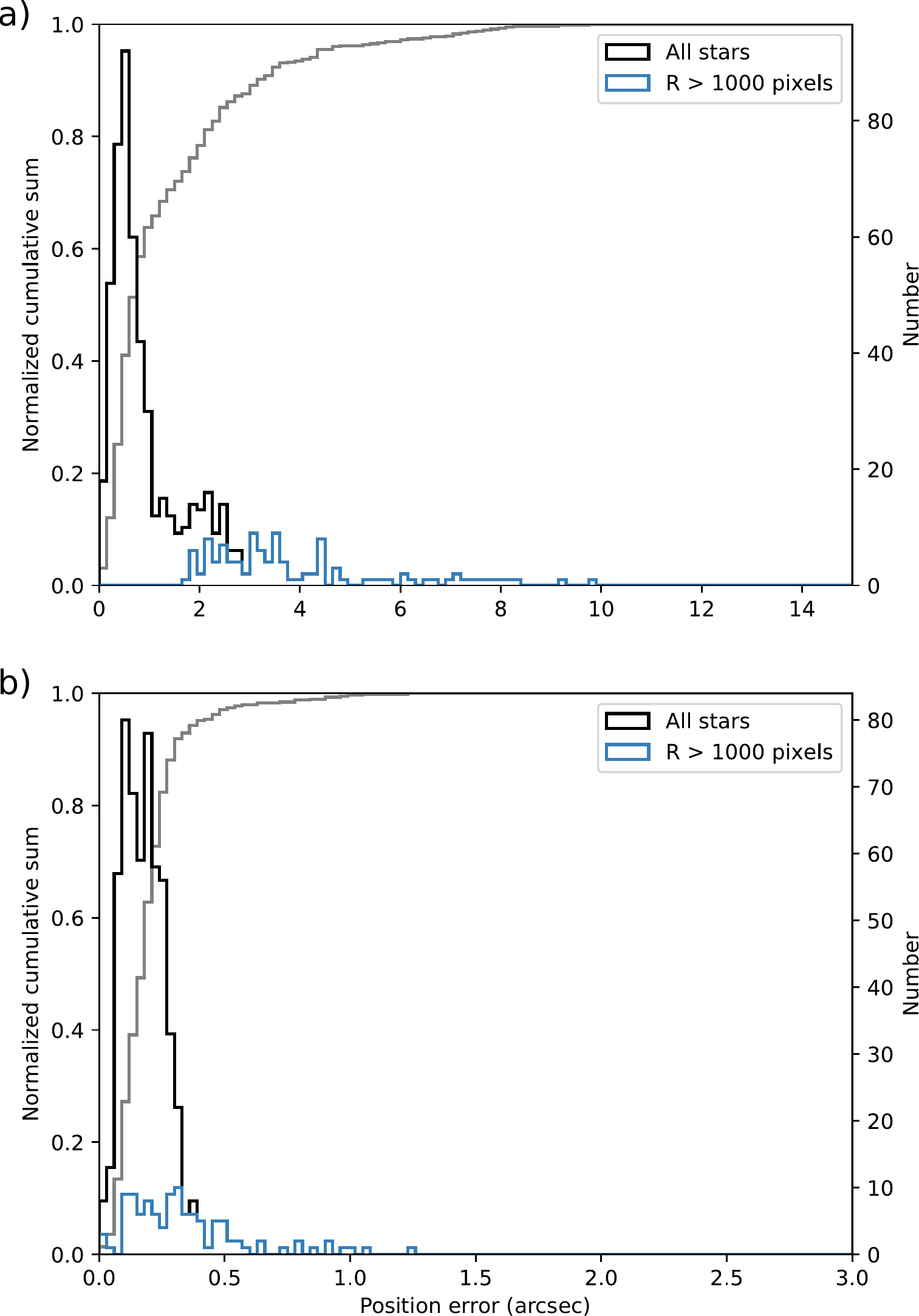}
  \caption{\textit{a)} Histogram of the position error after the first
    calibration level. The median of the error is around 2\,\arcsec
    and can be as large as 14\,\arcsec in the corners of the
    image. \textit{b)} Histogram of the position error after the
    second calibration level. The median of the error is now smaller
    than 0.3\,\arcsec which is more than 6 times better than the first
    level median error. The histograms of the stars in the corners,
    defined as having a radius with respect to the center of the field
    greater than 1000 pixels, i.e. 5.35\,\arcmin, are shown in
    blue. The cumulative histograms are shown in light grey.}
    \label{fig:astrom_hist_all}
\end{figure}

\begin{figure}
  \includegraphics[width=\columnwidth]{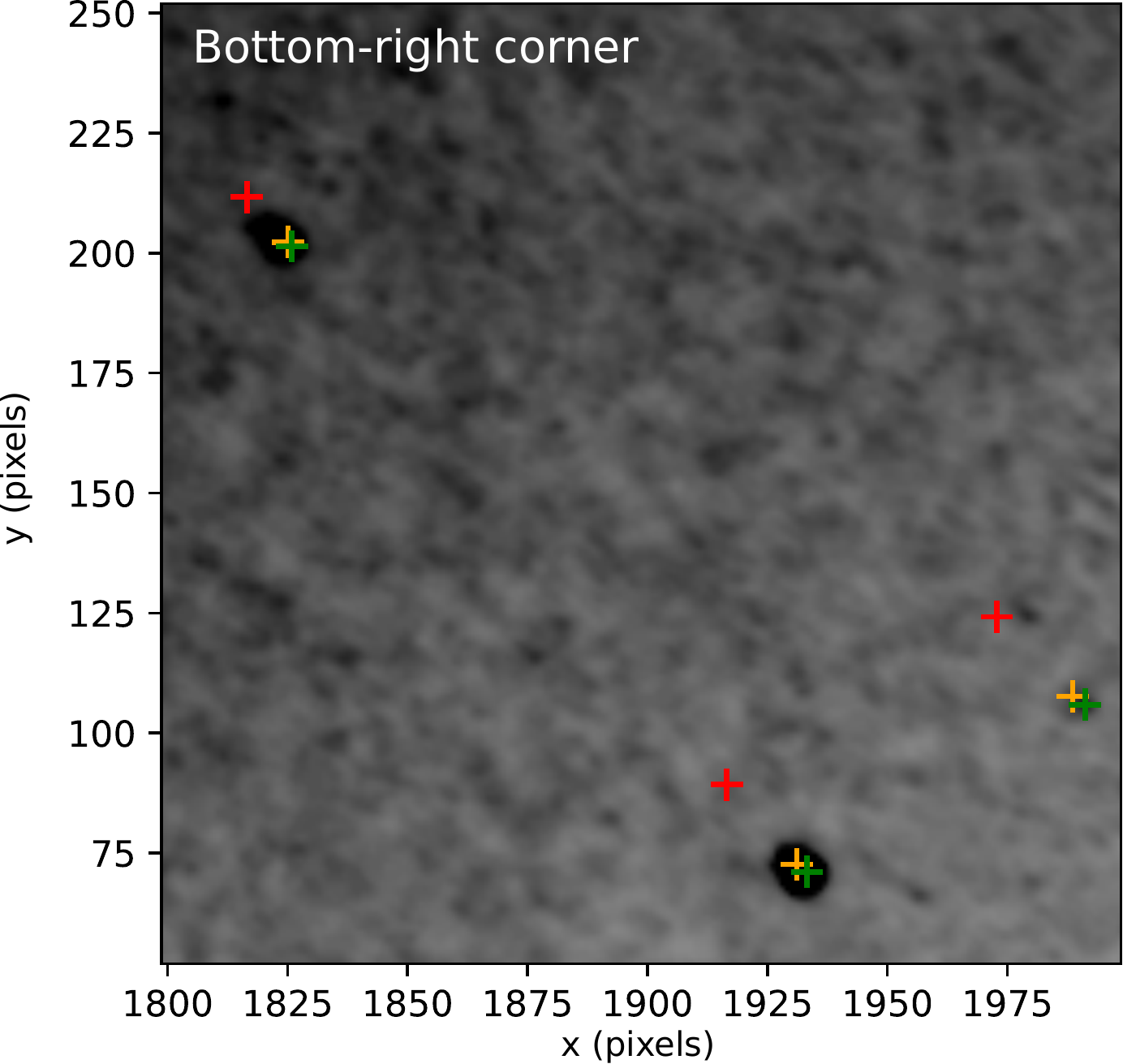}
  \caption{Position error in one region of the field of view of M\,31
    taken far from the center where the distortions are the most
    important. The crosses indicate the computed position of the stars
    after the first calibration step, in red, the second calibration
    step, in orange, and the third calibration step, in green.}
    \label{fig:astrom_im_all}
\end{figure}

\begin{figure}
  \includegraphics[width=\columnwidth]{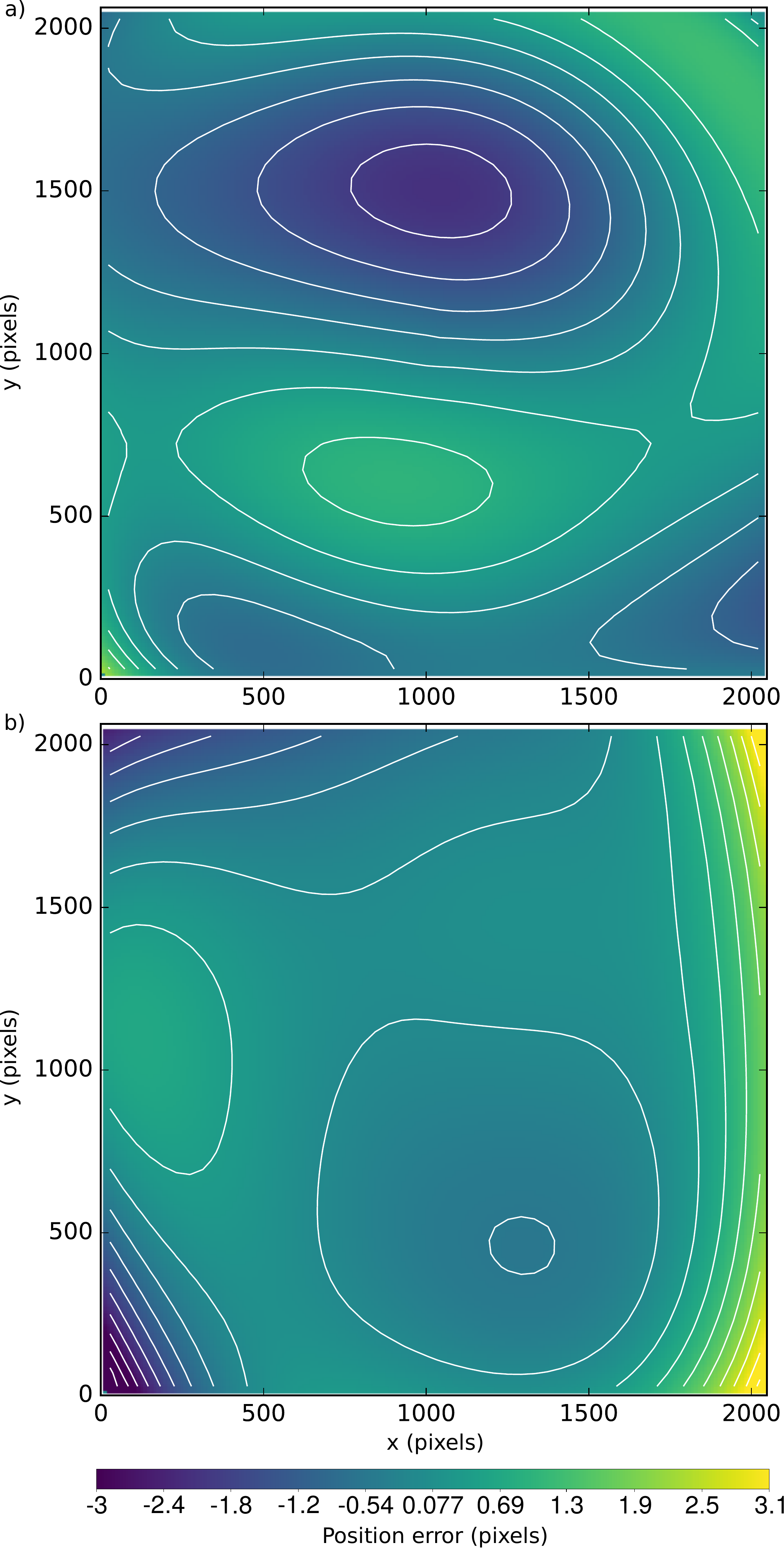}
  \caption{Residual distortion maps along the Y (\textit{a}) and X
    (\textit{b}) axes used for the third level of astrometric
    calibration. The position error is given in pixels. One pixel has
    a dimension of 0.32\,\arcsec.}
    \label{fig:dxdymaps}
\end{figure}

Even if the SIP model performs remarkably well at describing the
distortions of the field of view, the error can be as large as
1.5\,\arcsec, especially in the corners of the field where distortions
are the strongest (Fig.~\ref{fig:astrom_hist_all}). Primarily because
of SITELLE's optics which aberrations are radially dependant and
secondarily because of the image quality which is severely degraded in
the corners for reasons still unknown. To reduce this error down to
less than one pixel, we have no choice but to compute two distortion
maps, one for each axis (shown on Fig.~\ref{fig:dxdymaps}). The
interpolation of the displacement along both axes between the stars is
modelled with Zernike polynomials which are well suited to model
wavefront distortions.

The resulting calibration must have a precision better than one pixel
(0.32\,\arcsec), everywhere in the field. It seems difficult to
compare with other catalogues since the catalogue we use (Gaia DR1)
is, by far, the one which contains the more stars with the highest
precision. A comparison of our catalogue of emission-line point-like
sources (see section~\ref{sec:catalogue}) with another catalogue is
made in section~\ref{sec:comparison_with_other_catalogues}. It
suggests an upper-limit uncertainty of 0.21\,\arcsec on our
calibration which supports our calibration.

\subsection{Flux calibration}

\begin{figure}
  \includegraphics[width=\columnwidth]{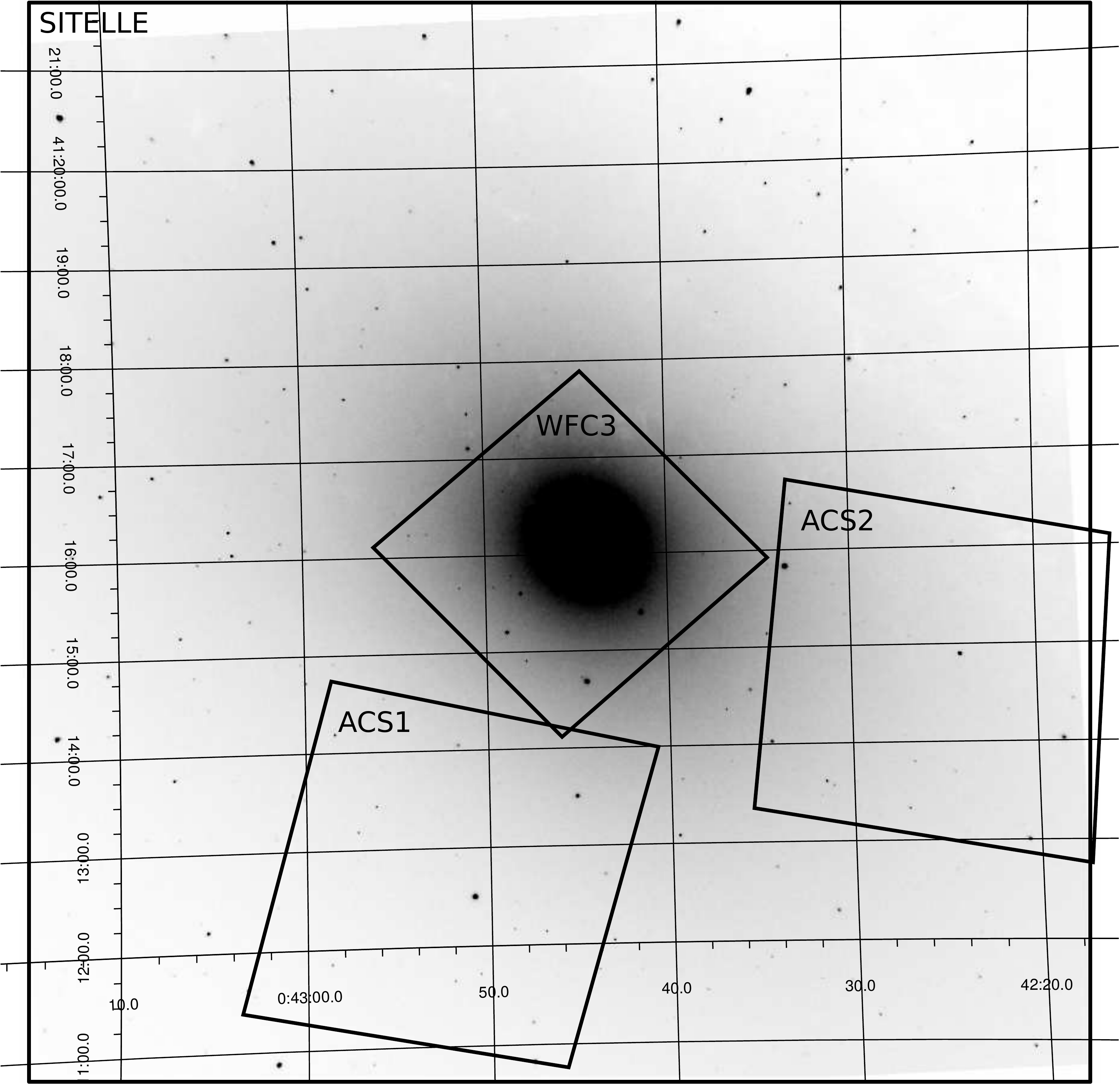}
  \caption{Positions of the HST fields over SITELLE's field of view
    used to refine the flux calibration.}
    \label{fig:flux_regions}
\end{figure}

\begin{figure}
  \includegraphics[width=\columnwidth]{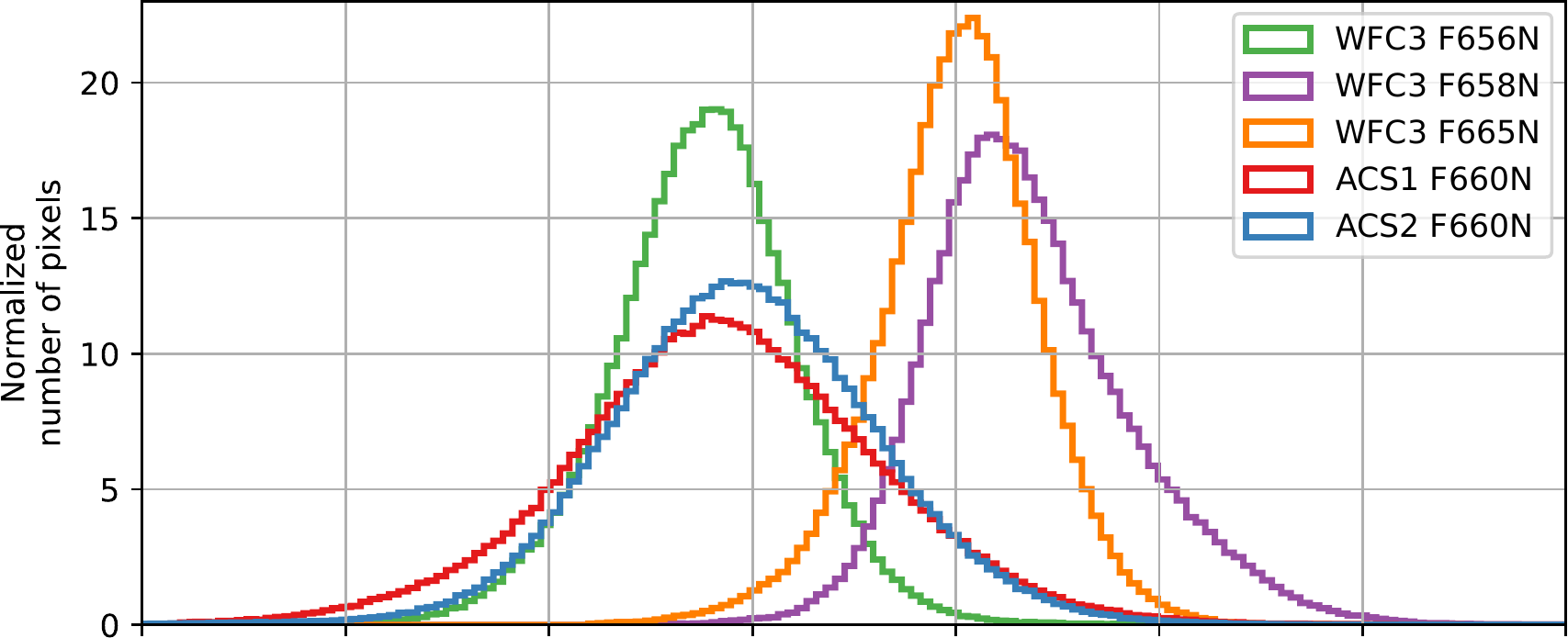}
  \caption{Histograms of the flux ratio between the integrated frames
    (which simulate HST frames) and the real HST frames. The median
    and standard deviation of theses distributions are reported in
    Table~\ref{tab:flux_calib}}
    \label{fig:hist_flux_all}
\end{figure}

The flux calibration was performed from two calibration sources: 1)
The spectrum of the spectrophotometric standard star GD71
\citep{Bohlin2003}, obtained in January 2016, which is used to
eliminate as much as possible any strong wavelength dependence ; 2)
the median combination of a set of 10 images of the standard star
HZ\,4 \citep{Bohlin2001} obtained right after the end of the cube
observation with photometric conditions similar to the observation
conditions. The exact value of the interferometer's modulation
efficiency (ME), which acts essentially as an additional throughput
loss, is the major source of uncertainty on the absolute flux
calibration \citet{Martin2016b}. Interferometric images of the laser
source have been obtained before and after the observation of the
target in order to measure the variation of ME at the calibration
laser wavelength (543.5\,nm) with respect to its nominal value
(85\,\%). We have measured a loss of 11.7\,\%. The initial flux
calibration of M\,31 in the SN3 filter has been corrected for this
loss (thus multiplied by a factor 1.13). But given the possible
uncertainty on this estimate, we have double-checked it using Hubble
Space Telescope (HST) narrowband images of the target. The advantage
of HST's narrowband filters is that they can easily be simulated by
integrating the spectral cube over the filter's well-known
transmission curve.  As SITELLE's cube flux is expressed in
erg\,cm$^{-2}$\,s$^{-1}$\,\AA$^{-1}$ and given the filter transmission
curve $F(\sigma)$, the integrated flux $\tilde{\phi}_F$, expressed in
erg\,cm$^{-2}$\,s$^{-1}$\,\AA$^{-1}$, is
\begin{equation}
  \label{eq:simulation_flux}
  \tilde{\phi}_F = \frac{\int_{-\infty}^{+\infty}\phi(\sigma)F(\sigma)d\sigma}{\int_{-\infty}^{+\infty}F(\sigma)d\sigma}\;.
\end{equation}
If one converts the flux in terms of surface brightness, expressed in
erg\,cm$^{-2}$\,s$^{-1}$\,\AA$^{-1}$ per HST pixel surface
($S_\text{HSTCamera}$), i.e.
\begin{equation}
  \tilde{B}_F = \frac{\tilde{\phi}_F }{S_\text{SITELLE}}S_\text{HSTCamera}\;,
\end{equation}
with $S_\text{SITELLE} = 0.32^2$\,arcsec$^2$, the image $\tilde{B}_F$,
once properly aligned, can be directly compared with the HST frame. We
have made this comparison for three different regions of the field of
view (called WFC3, ACS1, ACS2, shown in Fig.~\ref{fig:flux_regions})
and four different filters (F656N, F658N, F665N, F660N). Histograms of
the flux ratio between the integrated frames and the HST frames are
presented in Fig.~\ref{fig:hist_flux_all} and the first moments of
their distributions are listed in Table~\ref{tab:flux_calib}.

\subsubsection{Relative flux calibration}

We can estimate the pixel-to-pixel relative uncertainty from the flux
ratio maps. But we must be careful with the absolute calibration of
the Hubble images. Indeed, the average sky background intensity
reported in the WFC3 Instrument Handbook is $3.8\times 10^{-18}$
erg\,cm$^{-2}$\,s$^{-1}$\,\AA$^{-1}$\,arcsec$^{-2}$ at 6000\,\AA{}
that must be compared to the surface brightness of M\,31 which ranges
in our field of view from $5\times10^{-17}$ to
$5\times 10^{-15}$erg\,cm$^{-2}$\,s$^{-1}$\,\AA$^{-1}$\,arcsec$^{-2}$
accounting for 1.4\,\% in the WFC3 field and around 7\,\% in the ACS
fields.  Also, from the WFC3 Instrument Handbook the quoted photometry
accuracy is around 2--3\%. This uncertainty alone is sufficient to
explain the difference in the median ratio in the different fields
especially the 7\,\% difference between the F656N filter and the F658N
and F665N filters obtained in the same region at the center of the
field. It is thus difficult to assert that there is no gradient in the
absolute flux calibration smaller than 8\,\%. In the ACS fields a
higher limit of 4\,\% on the relative calibration can be deduced from
the standard deviation of the ratio histograms. In the central region
the smaller standard deviation is around 2\,\% which sets the higher
limit on the relative flux calibration in this region.

\subsubsection{Absolute flux calibration}

To get the most conservative estimate on the absolute flux calibration
correction we have computed the mean of the 5 median values of the
flux ratios (considered with their uncertainty) to obtain a mean flux
ratio of $96.9\pm1.4$\,\% which has been used to correct the flux
calibration of our data.

\begin{table}
  \centering
  \caption{Median and standard deviation of the histograms of the flux ratios shown in Figure~\ref{fig:hist_flux_all}. $\lambda_{\text{mean}}$ is the pivot wavelength of the HST filters.}
  \label{tab:flux_calib}
  \begin{tabular}{ccccl}
    Instrument & Filter & $\lambda_\text{mean}$ & Field & Median [Std]\\
               &&(in nm)&&(in \%)\\
    \hline
    WFC3 UVIS1 & F656N &656.15& WFC3 & 93.8 [2.3]\\
    WFC3 UVIS1 & F658N &658.56& WFC3 & 101.5 [2.6]\\
    WFC3 UVIS1 & F665N &665.60& WFC3 & 100.1 [2.1]\\
    ACS WFC & F660N &659.95& ACS1 & 94.3 [4.0]\\
    ACS WFC & F660N &659.95& ACS2 & 94.7 [4.1]\\
  \end{tabular}
\end{table}

\section{Emission line sources near the center of M\,31}
\label{sec:catalogue}
While the main objective of the SN3 data cube was to study the diffuse
gas around M31's nucleus, we were delighted to see a very large number
of point-like sources appearing as we scanned the reduced cube. Since
the brightest of them could likely be used to assess the validity of
our calibrations, we put some efforts into the systematic detection
and characterization of point-like emission-line sources in this cube.
The following section thus describes the detection technique, the
catalogue resulting from this analysis and comparisons with previous
catalogues, as well as a discussion of some interesting sources.

\subsection{Source detection}
\label{sec:detection_of_the_sources}
\begin{figure}
  \includegraphics[width=\columnwidth]{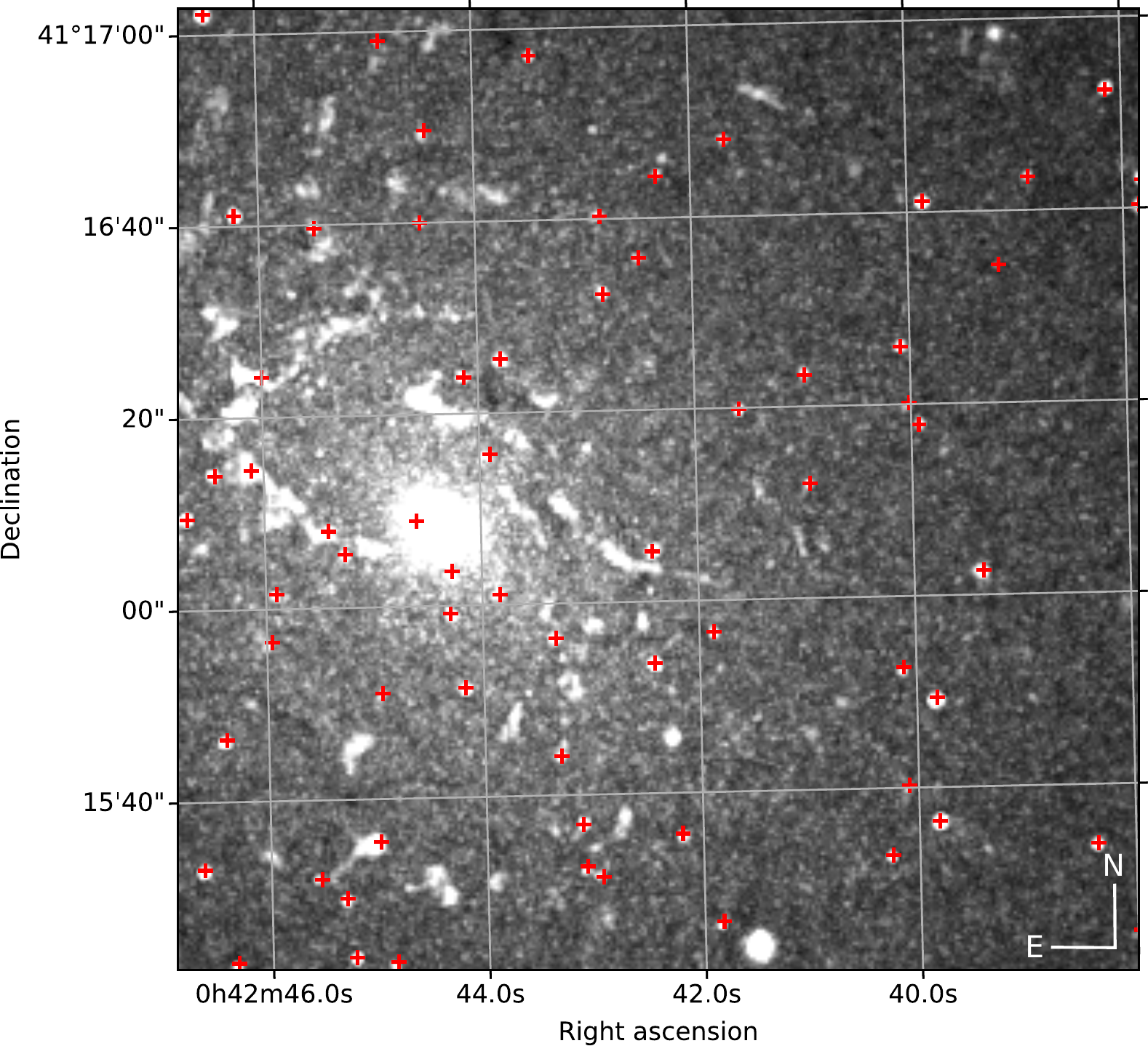}
  \caption{Cropped region of the detection frame around the center of
    M\,31. Red crosses indicate the positions of the sources of the
    catalogue.}
    \label{fig:det_frame1}
\end{figure}

The relatively large velocity range ($\sim 800$\,\kms{}) due to
galactic rotation near the center of M31 requires a dedicated
algorithm to detect emission-line sources. We devised such an
algorithm, which aims at detecting an excess of flux in at least one
spectral channel for a given pixel with respect to its surroundings.
First, the spectrum of this pixel is replaced by the median spectrum
of the 3$\times$3 pixels box centered on it (matching the mean seeing
during the observations). The median spectrum of a 9$\times$9 pixels
box (excluding the central 3$\times$3 pixels) is then subtracted.  The
maximum flux of this background-subtracted spectrum is then stored in
a detection map at the location of the pixel under investigation (see
Fig.~\ref{fig:det_frame1}).  This map, which therefore provides the
level of emission of each pixel above its surrounding background,
highlights numerous point sources, the brightest foreground stars and
some very bright portions of the filaments of diffuse gas around the
core of the galaxy. Finally we have checked this detection map by eye
for the presence of point-like sources which spectra have been
extracted and fitted manually.

\subsection{Extraction and fit of the sources}
\label{sec:extraction_and_fit_of_the_sources}
Once the identification of candidates was completed, the
backround-subtracted spectrum of each candidate has been integrated in
a 3$\times$3 pixels box around the central pixel. This time, the
background spectrum was chosen to be the median spectrum of a
30$\times$30 pixels box (excluding the central 3$\times$3 box), which
was found to be the optimal setting to maximize the signal-to-noise in
the whole galaxy though it might not be optimal near the very center
of M\,31 where there is an important background gradient.

\begin{figure}
  \includegraphics[width=\columnwidth]{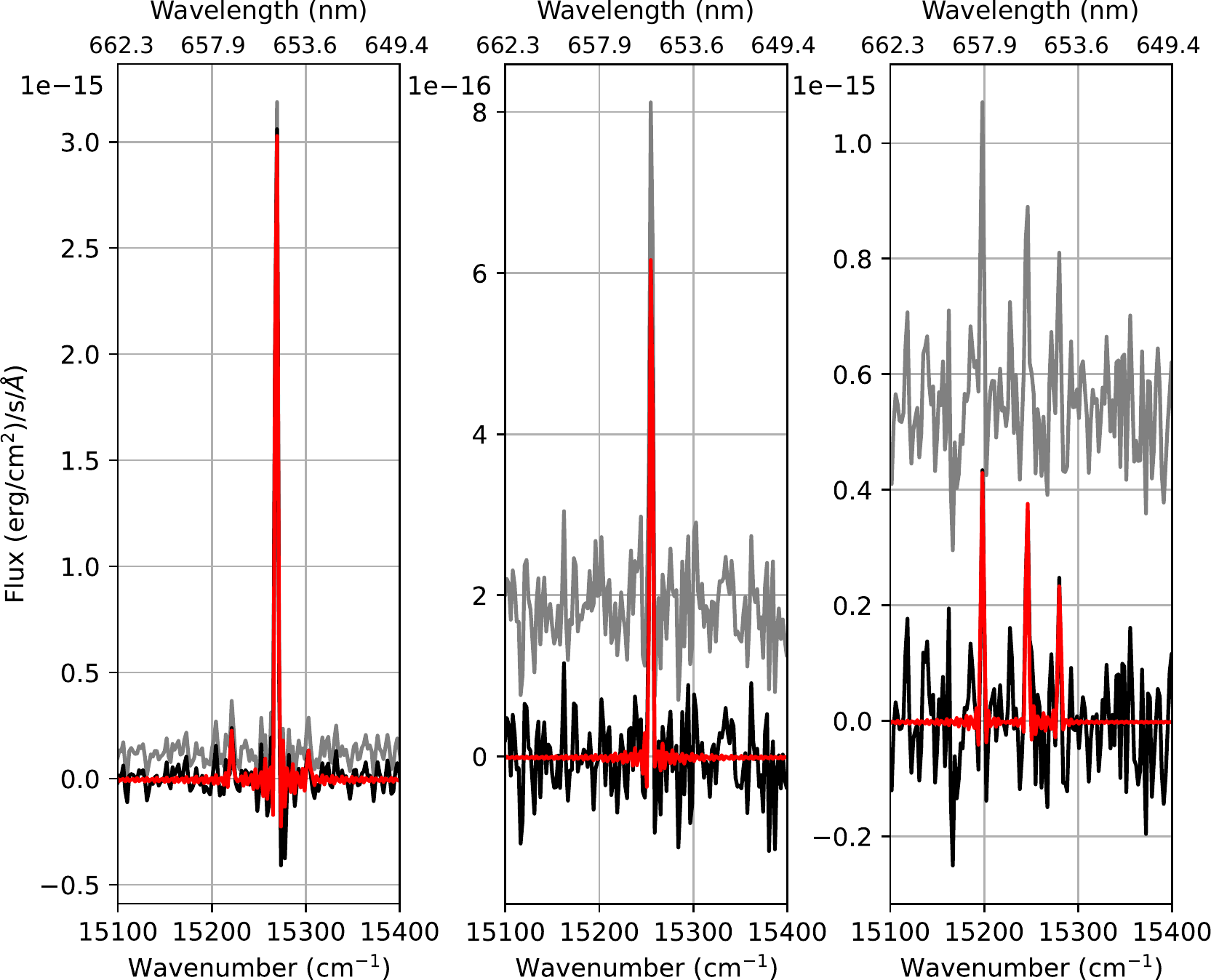}
  \caption{Examples of extracted spectra. The raw spectrum is in gray, the background corrected spectrum in black and the fit in red. Only the region around the \Ha{} line is shown.}
    \label{fig:fit_all}
\end{figure}

The spectrum was then fitted with a model made of 5 emission lines
(\Ha{}, \NII{}\,$\lambda 6548$, \NII{}\,$\lambda 6584$,
\SII{}\,$\lambda 6717$, \SII{}\,$\lambda 6731$) plus a flat
background. Note that the HeI\,$\lambda 6678$ line was not detected
anywhere. The analytic model describing the line spread function of
the emission lines is the convolution of a sinc, the instrumental line
shape, and a Gaussian, an approximation of the real line shape (see
Fig.~\ref{fig:fit_all}). The line model, described in details in
\citet{Martin2016}, has 3 varying parameters: amplitude, velocity and
broadening of the Gaussian; the FWHM of the sinc line is known and
fixed as it depends only on the maximum OPD of the interferogram (see
equation~5 of \citealt{Martin2016}). All the 5 emission lines were
considered to share the same velocity and the same broadening.

We have chosen to integrate the source spectrum into a 3$\times$3 box
in order to maximize the signal-to-noise ratio of our spectrum. But
the PSF clearly extends beyond this limit, so the flux must be
corrected for this aperture loss.  We have simulated our integration
model on a synthetic star with a Moffat point spread function
\citep{MoffatA.F.J.1969}. The real center of the synthetic star was
randomly positioned, with a uniform distribution, in the central pixel
to mimic the fact that the real center of the integrated source can be
anywhere in the central pixel of the 3$\times$3 box. The shape
parameters of the synthetic star (width and the Moffat parameter
$\beta$) were taken from a fit of multiple stars in the deep frame of
the cube (the deep frame represents the integral of the cube along the
spectral axis). From the analysis of 100\,000 trials we can conclude
that the measured flux must be multiplied by 1.88\,$\pm 0.07$.

The final correction made on the measured flux is the product of the
modulation efficiency loss, the flux calibration obtained from the
comparison with Hubble images and the computed PSF loss which gives a
final correction factor $\text{FC}$:
\begin{equation}
  \text{FC} = 1.13\, \times \frac{1}{0.969\pm0.014}\, \times 1.88\pm0.07 = 2.199\pm0.088 \;.
\end{equation}
The systematic relative uncertainty made on the measured flux,
independently of the uncertainty due to the SNR of the fitted spectra,
is thus 4\,\%.

\subsection{Description of the catalogue}

We have detected and catalogued a total of 797 \Ha{} emitting sources
in our field of view (see Table~\ref{tab:catalog_example}). The radial
velocity and the broadening of the lines as well as the flux of \Ha{},
\NII{}\,$\lambda 6584$, and the combined flux of
\SII{}\,$\lambda 6717$ and \SII{}\,$\lambda 6731$ have been reported
in the catalogue with their corresponding uncertainty. When the SNR of
a line was smaller than 5 and thus not clearly detected, we have
reported an estimation of the upper limit of the flux. This upper
limit was considered to be the minimum value between the estimated
amplitude plus 3 times its uncertainty or 5 times the uncertainty. As
the lines are modeled with the same velocity, the \SII{} and \NII{}
lines are always present in the fit results but the amplitude
uncertainty might be much higher than its measured value (as already
mentioned a line was considered detected for a SNR higher than 5). In
this case, the uncertainty used to calculate the upper limit on the
\SII{} or the \NII{} flux is strongly related to the background noise.

Examples of spectra are shown in Figure~\ref{fig:fit_all}. The
reported SNR is the maximum intensity attained by the spectrum
(i.e. by the \Ha{} or \NII{}\,$\lambda 6584$ line) corrected for the
background and the median intensity and divided by the standard
deviation of the residual of the fit (see
section~\ref{sec:extraction_and_fit_of_the_sources}).

We have cross-matched our sources with the catalogues of
\citet{Halliday2006} and \citet{Merrett2006}, aimed at the detection
and characterization of planetary nebulae specifically targeting the
\OIII{}\,$\lambda 5007$ line, and reported the corresponding ID in
Table~\ref{tab:catalog_example}. We have detected 304 of the 332
sources detected in \OIII{} by \citet{Merrett2006} (a 91.5\%
completeness) and 149 of the 154 sources observed by
\citet{Halliday2006} a 96.8\% completeness). A comparison of our
measured velocity with the velocity reported by \citet{Halliday2006}
is made in section~\ref{sec:comparison_with_other_catalogues}.

Supernovae remnants and novae were detected but have been removed from
the catalogue, the first ones because they are not point-like and the
second ones because of their transient nature.  They are discussed
independently in section~\ref{sec:objects_not_listed}.

\subsubsection{Luminosity function and completness}
\begin{figure}
  \includegraphics[width=\columnwidth]{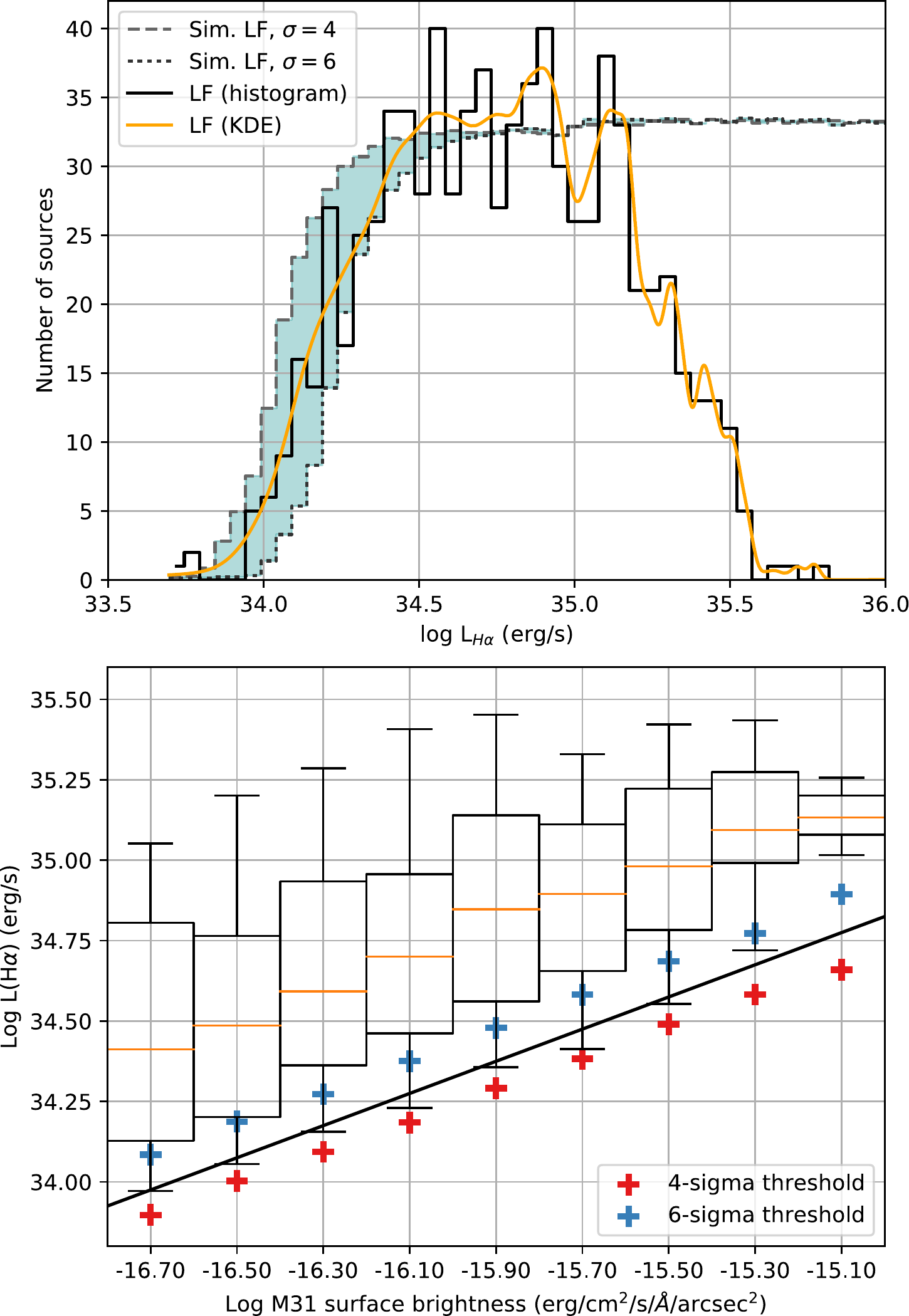}
  \caption{\textit{Top:} \Ha{} luminosity function $L_{H\alpha}$ of
    the catalogued sources, uncorrected for extinction. The kernel
    density estimation \citep{Rosenblatt1956, Parzen1962} of the
    luminosity function is also plotted as an orange line. The
    histograms simulated for a detection threshold at $4\sigma$ and
    $6\sigma$ are shown. The blue region between both simulated
    histograms indicates where the luminosity function should fall for a
    detection threshold between $4\sigma$ and
    $6\sigma$. \textit{Bottom:} Simulated 90\,\%-completness
    (luminosity under which the completness is smaller that 90\,\%)
    with respect to the background flux for a detection threshold at
    $4\sigma$ and $6\sigma$. The function from
    equation~\ref{eq:completness} is drawn in solid black. The box and
    whiskers plot represents the distribution of the luminosity of the
    catalogued sources per bins of the underlying galaxy's surface
    brightness. Whiskers are set at the 5th and 95th percentiles of
    the distributions.}
  \label{fig:lumf_all}
\end{figure}

The luminosity function, uncorrected for extinction, of the sources in
the catalogue is shown in Figure~\ref{fig:lumf_all} and has been
calculated with the relation
\begin{equation}
  \label{eq:lumf}
  L_{H\alpha} =  4 \pi D^2 F(H\alpha)\;,
\end{equation}
$D$ being the distance to the center of M\,31 assumed to be 0.783\,Mpc
\citep{Mochejska2000}.
Note that, from \citep{Azimlu2011, Walterbos1992} maximum luminosity
of planetary nebulae is $\sim 5\;10^{35}$\,erg\,s$^{-1}$ which is the
case of 99.6\,\% of our sources. Our detectability limit in terms of
flux $F(H\alpha)$ is around $10^{-16}$\,erg\,s$^{-1}$ with 4.1 hours
of integration over the very bright background of M\,31 which is
certainly one of the worst case scenario for a Fourier transform
spectrometer.

We have verified the completeness of our sample by simulating the
detection of 10 million randomly positioned emission-line
point-sources in one channel of a spectral cube. Considering only the
photon noise contribution, the noise is the same in every channel and
its amplitude is the square root of the total number of counts
accumulated in each pixel
\citep{Drissen2012,Drissen2014,Maillard2013}. Starting with a map of
the total number of counts in each pixel we can add a group of
randomly positioned sources with a log-uniform flux distribution and
compute its square root to obtain a map of the photon noise. Each
source has its flux concentrated in only one pixel and one channel
which greatly simplify the simulation of the detection process. The
noise is multiplied by $\sqrt{9}$ to simulate the 3$\times$3 binning
used for the detection and the extraction of the spectra. The flux of
the one-pixel star is divided by 1.88 to simulate the loss of flux in
the wings of the point spread function. It is also divided by 1.25 to
simulate the fact that its real flux is distributed as a sinc line
spread function with a width of 1.25 channels \footnote{We recall that
  the flux $F$ of a sinc line of amplitude $A$ and width $W$ is
  $F = AW$}. All the pixels showing a value above the background level
plus a given number of times the calculated noise $\sigma$ are
considered as detected sources. Note that the 10 millions sources were
simulated by sets of a thousand sources in a SITELLE frame of 4
million pixels. So that the probability of having multiple sources at
the same pixel is negligible. The histograms of the detected sources
for a threshold at 4$\sigma$ and 6$\sigma$ are shown in
Figure~\ref{fig:lumf_all}. They reproduce well the slope of the
luminosity function at low luminosity level, confirming that the
completeness is good at a detection threshold of $\sim 5\sigma$.

In terms of homogeneity of the completness in the field of view, it is
clear that the level of detectability changes with the surface
brightness of M\,31. We have used the results of the simulation
detailed above to compute the 90\,\%-completness (i.e. the limit
luminosity under which the completness falls under 90\,\%) with
respect to M\,31's surface brightness $B$ for a detection
threshold of $4\sigma$ and $6\sigma$ (see
Fig.~\ref{fig:lumf_all}). The function which best fits the mean of the
simulated data is
\begin{equation}
  \label{eq:completness}
  \log(L_{\text{H$\alpha$}}) = 0.5 \log(B) + 42.325\;.
\end{equation}
We have also reported in Figure~\ref{fig:lumf_all} a box and whisker
plot representing the distribution of the luminosity of the catalogued
sources per bins of the underlying galaxy's surface
brightness. Whiskers are set at the 5th and 95th percentiles of the
distributions. If we examine closely the whiskers corresponding to the
5th percentile of the luminosity distribution we can see that they
nearly all fall between the $4\sigma$ and $6\sigma$ simulations,
except for the highest brightness bin which contains much less sources
than the others. This result confirms the adopted simulation process.

\subsubsection{LGGS cross-match}

\begin{table}
  \caption{Results of the detection of our sources in the LGGS ratio
    frames when they are considered together and independently (see
    text for details). Detection is said non-applicable (N/A) when the
    source is situated in the saturated region near the centre of
    M\,31 (in at least one of the frame when frames are considered
    together). A source is considered undetected in both frames if it
    is not detected in any of the frames and if it is not situated in
    a saturated region. A source is considered detected in both frames
    when it is detected in at least one of the frames.}
  \label{tab:lggs}
  \begin{tabular}{lccc}
    \hline
    LGGS ratio frame & detected & undetected & N/A\\
    \hline
    \Ha{}/R&533&147&117\\
    \OIII{}/B&389&376&32\\
    \Ha{}/R or \OIII{}/B&595&117&85\\
    \hline
  \end{tabular}
\end{table}

\begin{figure*}
  \includegraphics[width=\linewidth]{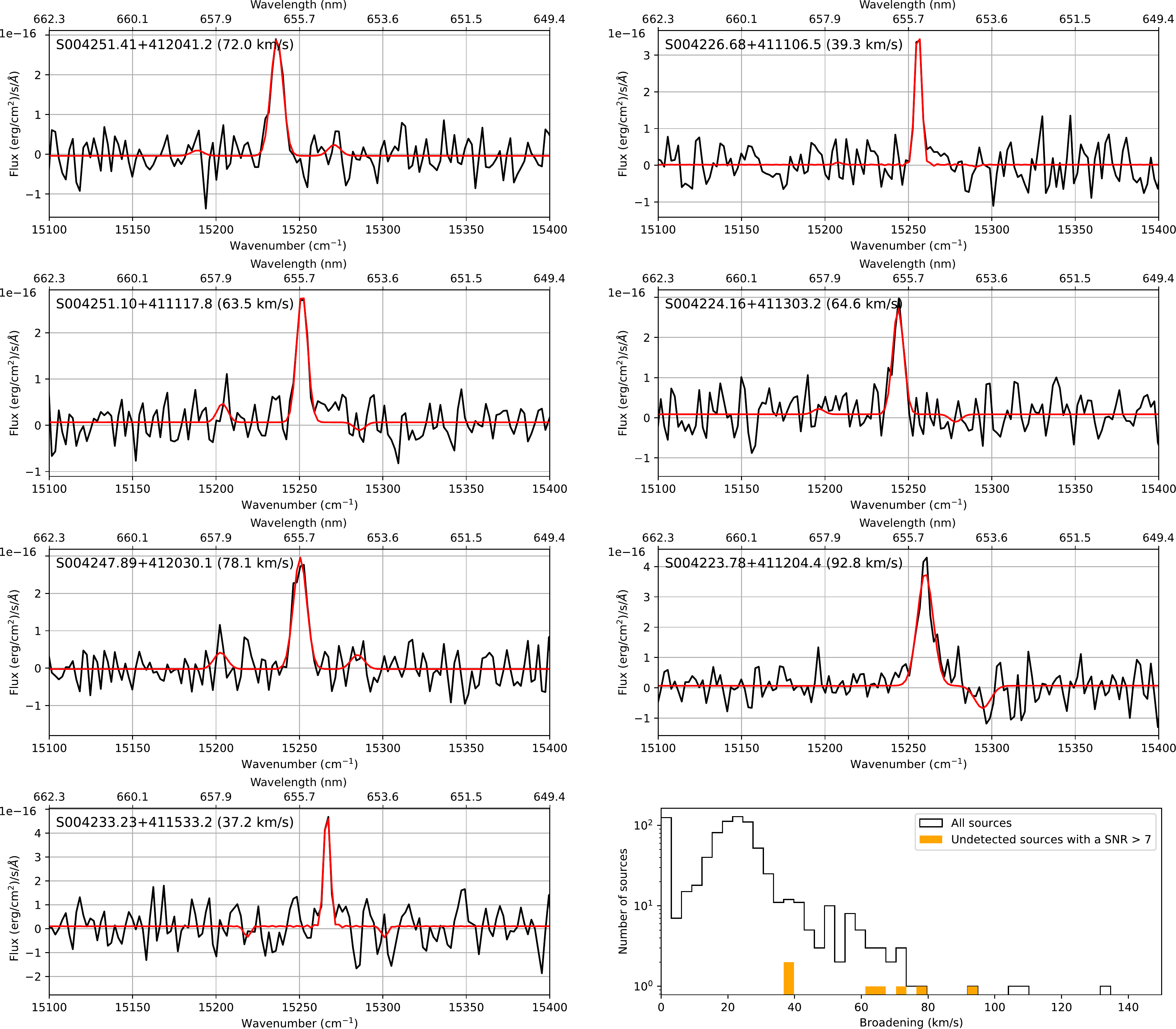}
  \caption{Spectra, cropped around the \Ha{} lines and the \NII{}
    doublet, of the seven sources having a SNR greater or equal to 7
    and undetected in the LGGS ratio images (see text for
    details). The fit of the 3 lines is shown in solid red. The source
    id, as well as the width of the \Ha{} line are indicated above
    each spectrum. At the bottom-right, we show the histogram of the
    line broadening of all the sources (black line) and the broadening
    of the seven displayed sources (orange bars). Two of the three
    objects with a broadening larger than 100 km/s have a split
    profile (see section~\ref{sec:split}). The apparent bimodality of
    the histogram (with two peaks at 22 and 0\,\kms) is an artefact of
    the fitting algorithm which can be reproduced when fitting low
    signal-to-noise ratio lines (see text for details).}
    \label{fig:lggs}
\end{figure*}

We have verified if our sources were also detectable in the images of
the Local Group Galaxies Survey (LGGS, \citealt{Massey2006}; field
F5). To do so, we have divided the narrowband images centered on \Ha{}
and \OIII\,$\lambda 5007$ by their corresponding broad-band images R
and B and checked by eye if a source could be seen at the location of
our emission-line sources.  Note that a region of 2.5 arc-minutes and
15 arc-seconds at the center of M\,31 in, respectively, the R and B
images is saturated and cannot be studied; sources in this region are
marked with ``N/A'' in the corresponding columns.  We report in
table~\ref{tab:lggs} the number of detected and undetected sources in
each ratio image. A total of 595 of our sources were detected in at
least one of the LGGS ratio images while 117 were clearly not detected
; 85 are not detected in one of the frame but are situated in a
saturated region in the other frame, in which case the non-detection
is less clear. If, in most cases, those undetected sources have a
relatively low SNR (which may explain the non-detection), with a
median of 5, seven of them have a SNR larger than 7 and should have
shown up in the LGGS ratio images. It indicates that they could
display new or variable emission lines (the field F5 of the LGGS data
has been obtained the 22th of September 2001). Another interesting
fact is that all those objects share a similar spectral morphology,
namely that they display a broad \Ha{} emission line (broader than the
instrumental resolution) with no sign of \NII{} (see
Fig.~\ref{fig:lggs}). From the histogram shown in Fig.~\ref{fig:lggs}
it appears that this broadening is significantly larger than that of
the vast majority of the catalogued sources.

The histogram of Figure~\ref{fig:lggs} shows an apparent bimodality
(with two peaks at 22 and 0\,\kms) which is an artefact of the fitting
algorithm that can be reproduced when fitting low signal-to-noise
ratio lines. The bimodality beeing clearly separated at 4\,\kms (with
no values of the broadening at 4\,\kms) all the measured broadening
values smaller than 4\,\kms have been removed from the catalogue.

\subsubsection{Split profiles}
\label{sec:split}
\begin{figure}
  \includegraphics[width=\columnwidth]{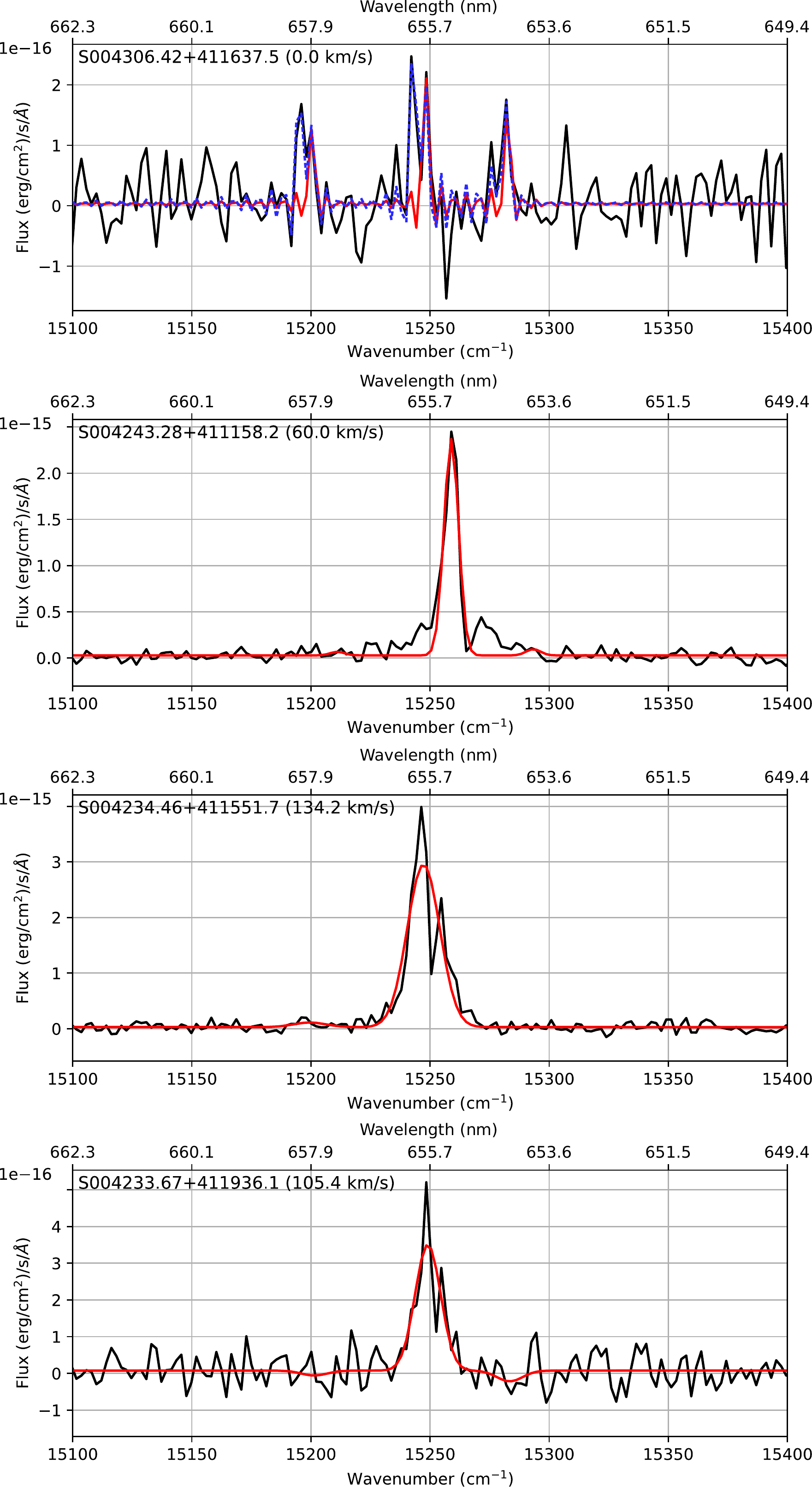}
  \caption{Spectra of sources showing split profile. The observed
    spectrum is displayed in black. We can clearly see the departure
    from the fitted model, in red. The measured broadening of the
    lines is reported in parenthesis. The top spectrum is clearly an
    expanding shell. The red-shifted and the blue-shifted components
    have been fitted together. The resulting fit is plotted in dotted
    blue. The measured expansion velocity is 90.7$\pm$1.4\,\kms{}. As
    it was fitted with a sinc model the broadening of the lines is, by
    definition, 0\,\kms{}.}
    \label{fig:pcyg}
\end{figure}

Four sources are displaying split profiles which are shown in
Figure~\ref{fig:pcyg}. They have been added the commentary ``Split''
in the catalogue. The first object (S004306.42+411637.5) is clearly an
expanding nebula with a velocity difference of 90.7$\pm$1.4\,\kms{}
between the two components. The second object (S004243.28+411158.2)
could also be seen as a single line with two satellites lines or a
broadened line with some absorption (like a PCygni profile). No \SII{}
lines are detected in its spectrum, and no stellar source is seen in
the deep SN3 image. The three others show a broad, split \Ha{} line,
with no evidence for forbidden lines. A relatively bright point source
is visible in the deep SN3 image in all cases.

\subsubsection{Elongated PSF}

A non negligible fraction of the outer portion of the field of view
shows an elongated point spread function. The origin of the elongation
is still unknown and could be related to an error in the model of the
telescope or the optics of SITELLE (Drissen et al. in
preparation). The 89 sources found in the affected region were marked
with the commentary ``elongated PSF''. As 90\,\% of the flux is
concentrated in the principal lobe of the PSF the effect on the
completeness is likely to be under 10--15\,\%. The computed flux is
also likely to be underestimated by the same amount. The measured
velocity suffers the same bias as the sky lines and as thus been
calibrated. No velocity bias should be observed.

\subsection{Comparison with the catalogue of \citet{Halliday2006}}
\label{sec:comparison_with_other_catalogues}

\defcitealias{Halliday2006}{Ha06}
\defcitealias{Merrett2006}{Me06}

\citet{Halliday2006} (henceforth \citetalias{Halliday2006}) have
measured the radial velocity of 723 planetary nebulae in M\,31, 154 of
which are present in the field observed with SITELLE. We have detected
148 of them (96\,\%) which is an excellent completeness rate
considering the fact that our detection is based on the \Ha{}
while theirs is based on the usually brighter \OIII{} line.

\begin{figure}
  \includegraphics[width=\columnwidth]{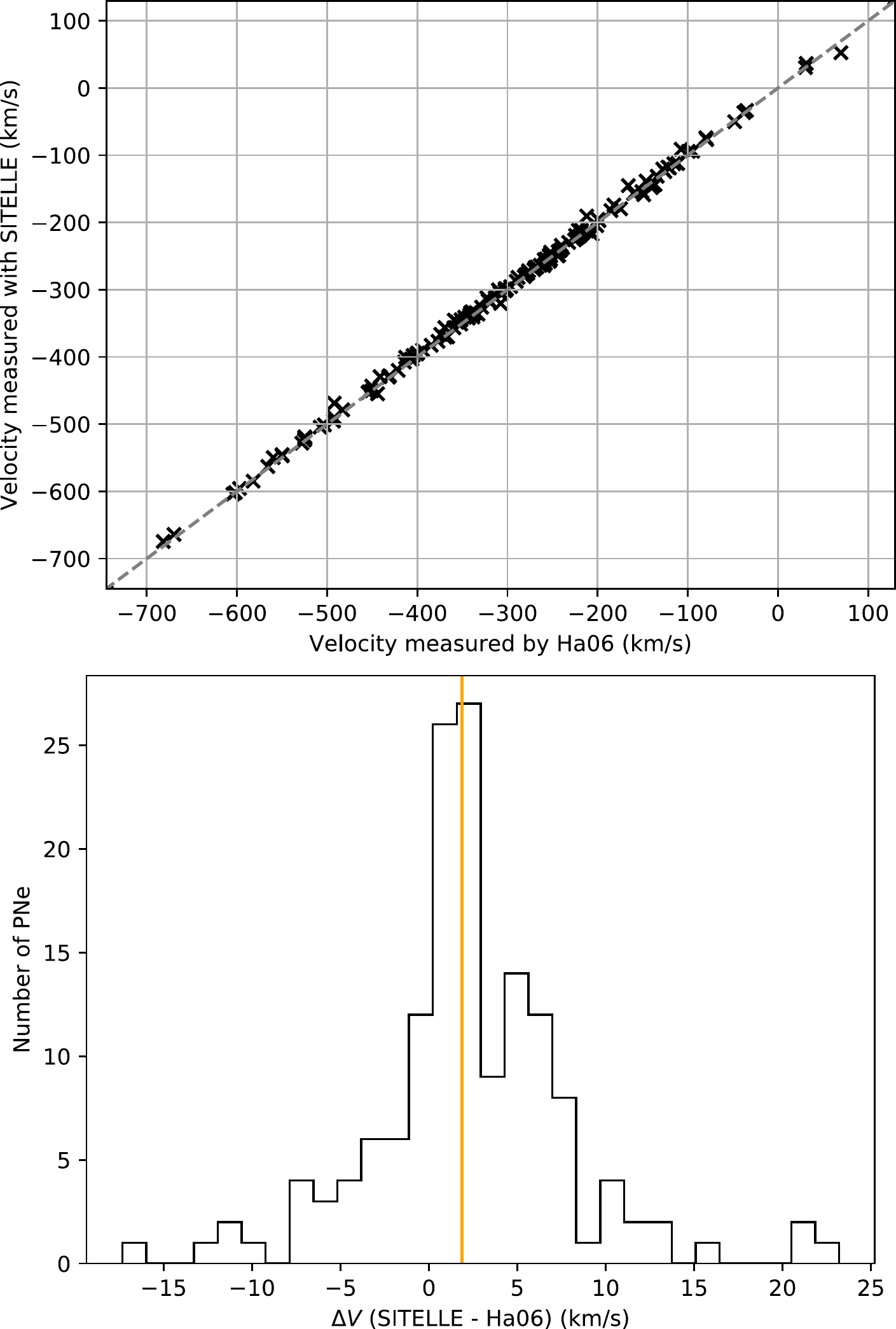}
  \caption{\textit{Top:} Velocity measured with SITELLE vs. velocity
    measured by \citetalias{Halliday2006}. The one-to-one line is
    shown in dotted grey. \textit{Bottom:} Histogram of the velocity
    difference with \citetalias{Halliday2006}. The median of the
    distribution is shown in orange.}
    \label{fig:stats_Hall1}
\end{figure}

\begin{figure}
  \includegraphics[width=\columnwidth]{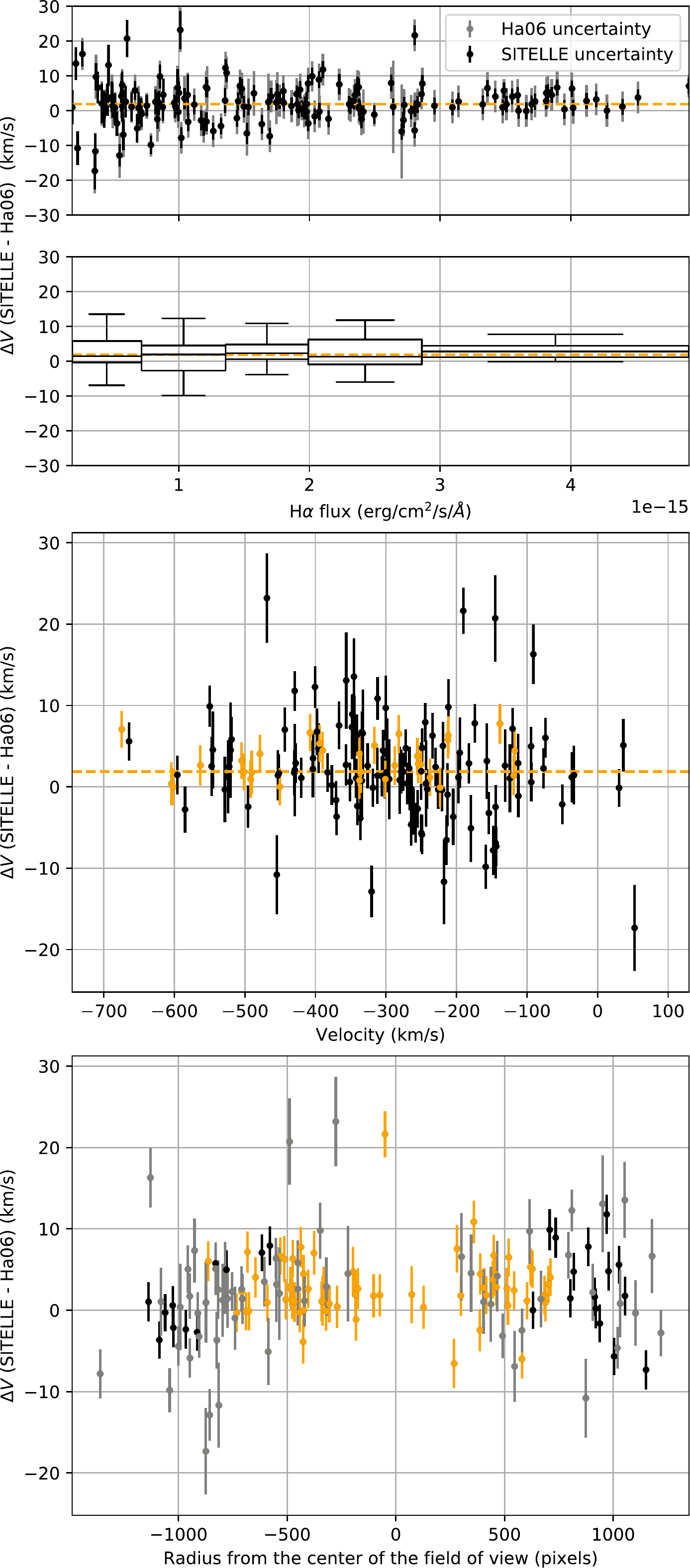}
  \caption{\textit{Top:} Velocity difference with
    \citetalias{Halliday2006} vs. \Ha{} flux. The box plot represents
    the distribution of the difference by bins of equal number of data
    points. \textit{Center:} Velocity difference with
    \citetalias{Halliday2006} vs. velocity. The 10\,\% most luminous
    sources are shown in orange \textit{Bottom:} Velocity difference
    with \citetalias{Halliday2006} vs. radius with respect to the
    center of the field of view (the radius is set negative when the
    distance along the declination axis, i.e. the y axis, is
    negative). The sources in orange are situated in the region where
    the sky lines velocity could not be measured. Their calibration
    thus heavily rely on the model described in
    section~\ref{sec:wave_calib_method}.}
    \label{fig:stats_Hall2}
\end{figure}

We have plotted their velocities against ours and found that they fall
very near the one-to-one line (Fig.~\ref{fig:stats_Hall1}). The
histogram of the velocity difference with Ha06 is also shown in
Fig.~\ref{fig:stats_Hall1}. The median difference is 1.8\,\kms{} and
the standard deviation is 6.1\,\kms{}. If we quadratically sum the
uncertainty reported by \citetalias{Halliday2006} and ours, the median
uncertainty on the velocity difference is 5.6 \kms{}. These results
confirms at the same time their calibration and ours as long as both
quoted uncertainties are correct. Because we could not have any
precise measurement of the OH lines velocity in the center of the
field of view, we have been using a model to estimate the velocity
correction in the center based on the values obtained on the borders
of field (see section~\ref{sec:wavelength_calibration} and
Fig.~\ref{fig:sky_map_all}). Any error on the model would translate
into a correlation between the velocity difference and the radius with
respect to the center of the image. This is not the case (see
Fig.~\ref{fig:stats_Hall2}). We have also checked that there was no
correlation with the velocity and that the difference was getting
smaller at higher flux (Fig.~\ref{fig:stats_Hall2}). Note that
\citetalias{Halliday2006} are measuring the \OIII{} lines while we are
providing the flux in \Ha{}; those two properties are only partially
correlated.

\begin{figure}
  \includegraphics[width=\columnwidth]{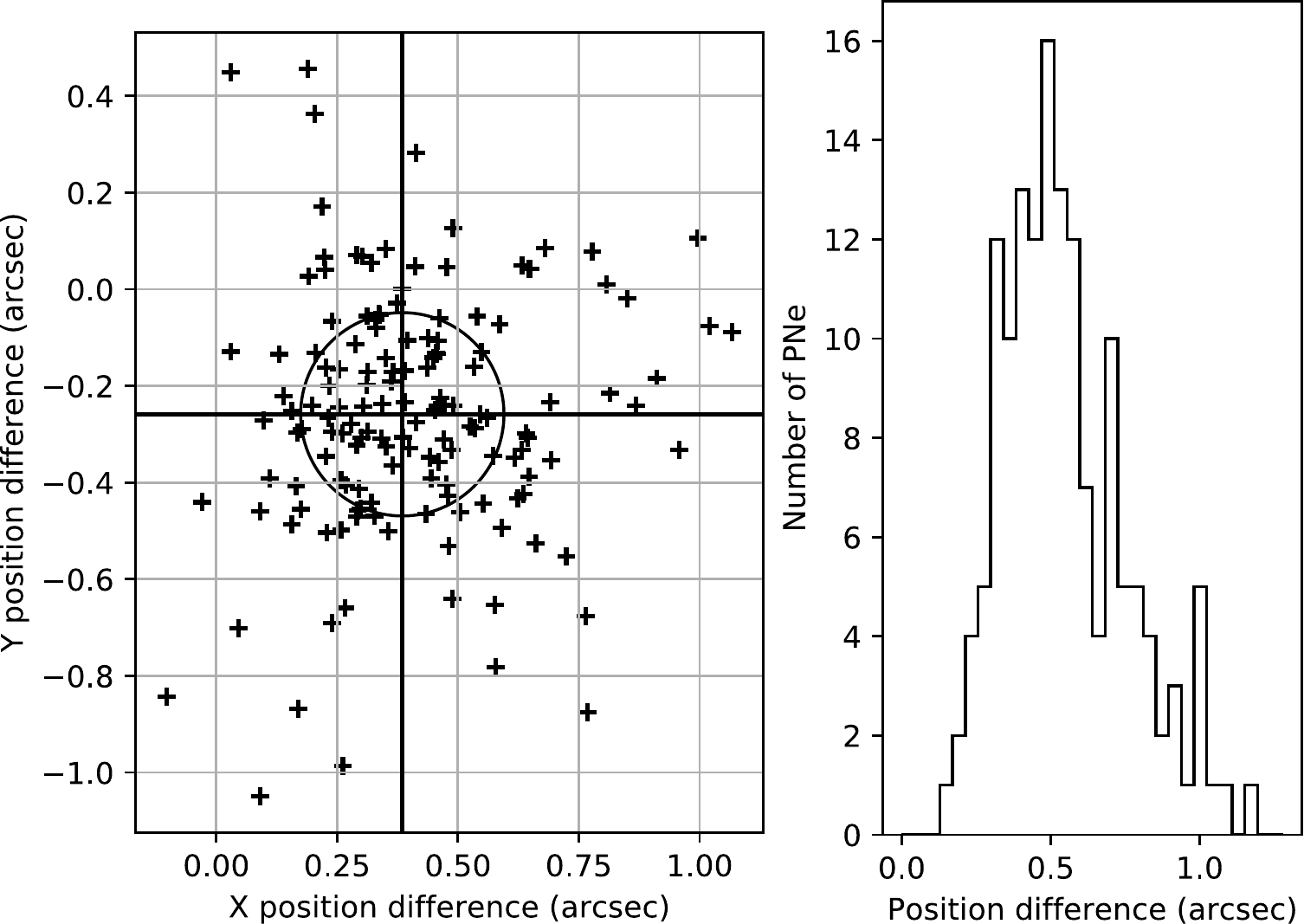}
  \caption{\textit{Left: } Scatter of the position differences along
    the X and Y axes of the image of the sources compared with
    \citetalias{Halliday2006}. The vertical and the horizontal black
    lines indicate the median of the differences in X and Y
    respectively. A circle represents the standard
    deviation. \textit{Right: } Histogram of the position differences
    with \citetalias{Halliday2006}.}
    \label{fig:stats_Hall3}
\end{figure}

When comparing the position of our sources with the position reported
by \citetalias{Halliday2006}, we find a median difference of
$0.46\,\arcsec$ (\citetalias{Halliday2006} quote an absolute
uncertainty of $1\,\arcsec$, with a relative uncertainty of
$0.4\,\arcsec$ on their positions), with a standard deviation of
$0.21\,\arcsec$, which is smaller than one SITELLE pixel and very
similar to our estimate of the uncertainty (see
Fig.~\ref{fig:stats_Hall3}).

The comparison with the catalogue of \citetalias{Halliday2006}
therefore confirms the quality of our wavelength and astrometric calibration.

\subsection{Comparison with the catalogue of \citet{Merrett2006}}

\begin{figure}
  \includegraphics[width=\columnwidth]{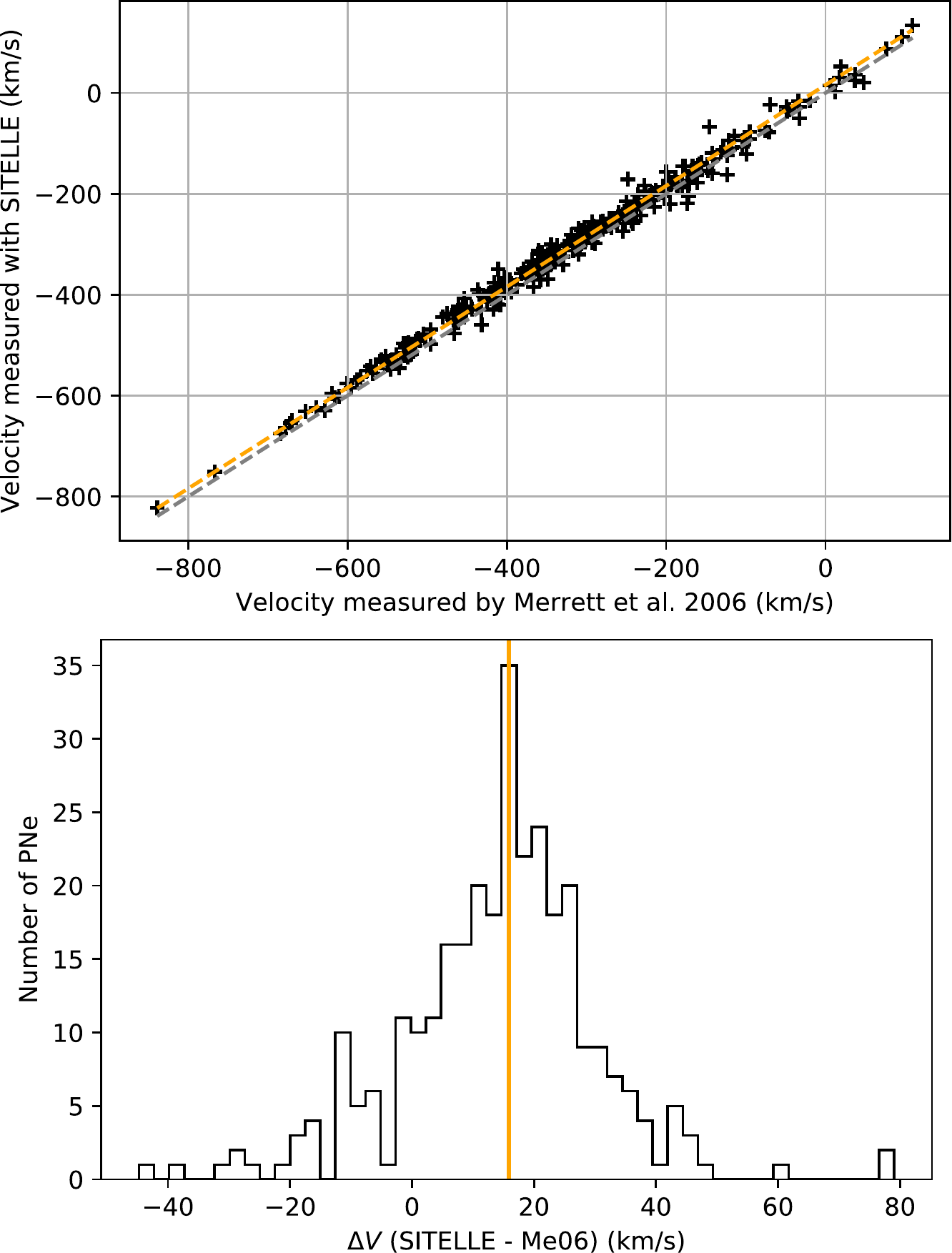}
  \caption{\textit{Top:} Velocity measured with SITELLE vs. velocity
    measured by \citetalias{Merrett2006}. The one-to-one line is shown
    in dotted grey. The same line corrected for the median of the
    distribution of the velocity difference (see bottom histogram) is
    shown in dotted orange. \textit{Bottom:} Histogram of the velocity
    difference with \citetalias{Merrett2006}. The median of the
    distribution is shown in orange.}
  \label{fig:vel_vs_merrett_all}
\end{figure}

\citet{Merrett2006} (henceforth \citetalias{Merrett2006}) have
measured the radial velocity of 2615 planetary nebulae in M\,31, 330
of which are present in the field observed with SITELLE. We have
detected 305 of them (92\,\%) which is a good completeness rate
considering the fact that our detection is based on the \Ha{} while
theirs is based on the usually brighter \OIII{} line. The quoted
uncertainty on their measurement is 17\,\kms{} with a systematic error
around 5--10\,\kms{} which is much higher than the uncertainty of
\citetalias{Halliday2006}'s catalog. The distribution of the velocity
difference with their catalogue is shown in
Figure~\ref{fig:vel_vs_merrett_all}. This distribution' standard
deviation of 16\,\kms{}  is strongly dominated by the uncertainty
of their measurement and confirms their estimation. The distribution
is also biased by 16\,\kms{} which is also consistent with the
comparison they made with the data of \citetalias{Halliday2006} (they
found a systematic difference of 17\,\kms{}).

We can further check the quality of our photometry by comparing the
flux ratio F(\OIII{})/F(\Ha{} + \NII{}) with respect to the relative
m$_{5007}$ magnitude reported in their catalogue. Note that the \OIII
flux also comes from their value of $m_{5007}$ since
\citetalias{Merrett2006}:
\begin{equation}
  m_{5007} = -2.5 \log\left(F(\OIII)\right) - 13.74\;.
\end{equation}
\begin{figure}
  \includegraphics[width=\columnwidth]{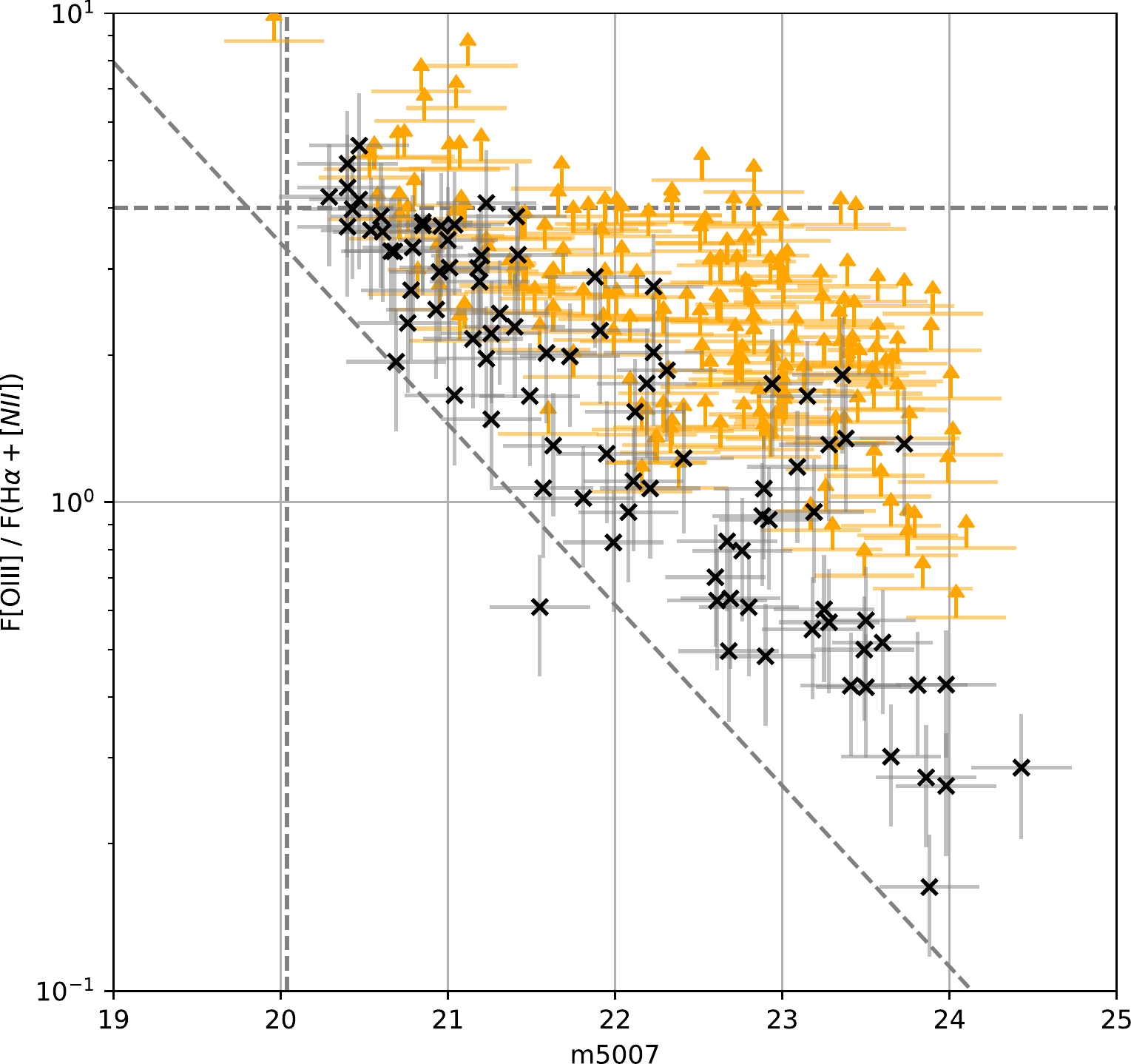}
  \caption{Flux ratio F(\OIII{})/F(\Ha{} + \NII{}) with respect to the
    relative m$_{5007}$ magnitude from \citetalias{Merrett2006}. The
    sources which \NII{} line flux could only be given an upper limit
    are plotted in orange. The other are plotted with black
    crosses. The error bars are also shown. The region in this graph
    where all the planetary nebulae are supposed to be is defined by
    the 3 dotted gray lines. The vertical line corresponds to the
    planetary nebulae luminosity function zero-point quoted by
    \citet{Ciardullo2002} and the two others are defined in
    \citet{Herrmann2008}.}
  \label{fig:m31_oiii_ha_nii_vs_m5007}
\end{figure}
We can see that most of our common sources fall in the wedge-like zone
where we can find all the planetary nebulae of the Local Group
Galaxies M31 and M33 \citep{Schonberner2010}. This zone is located
between the lines defined by \citet{Herrmann2008}:
\begin{equation}
  \log\left(\frac{F(\text{\OIII})}{F(\text{\Ha} + \text{\NII})}\right) = -0.37 M(\OIII) - 1.16\;,
\end{equation}
and
\begin{equation}
  \log\left(\frac{F(\text{\OIII})}{F(\text{\Ha} + \text{\NII})}\right) \simeq 4\;,
\end{equation}
where $M(\OIII)$ is the absolute magnitude in the \OIII$\lambda5007$
line which has been computed from the relation
\begin{equation}
  M(\OIII) = m_{5007} - A_V - \mu\;,
\end{equation}
considering the values of the reddening $E(B-V) = 0.062$\,mag
\citep{Schlegel1997} and a distance $D$ to M\,31 of 750\,kpc
\citep{Freedman2000} which translate to an interstellar extinction
$A_V = 3.1 E(B-V) = 0.19$\,mag and a distance modulus
$\mu = 5\log(D) - 5 = 24.38$\,mag. We have chosen here the same references as
\citet{Ciardullo2002} to keep its value of the planetary nebulae
luminosity function zero-point $M_{\star} = -4.53$ mag which is also
reported in Figure~\ref{fig:m31_oiii_ha_nii_vs_m5007}.

\subsection{Objects not listed in the catalogue}
\label{sec:objects_not_listed}
\subsubsection{Novae}

\begin{figure}
  \includegraphics[width=\columnwidth]{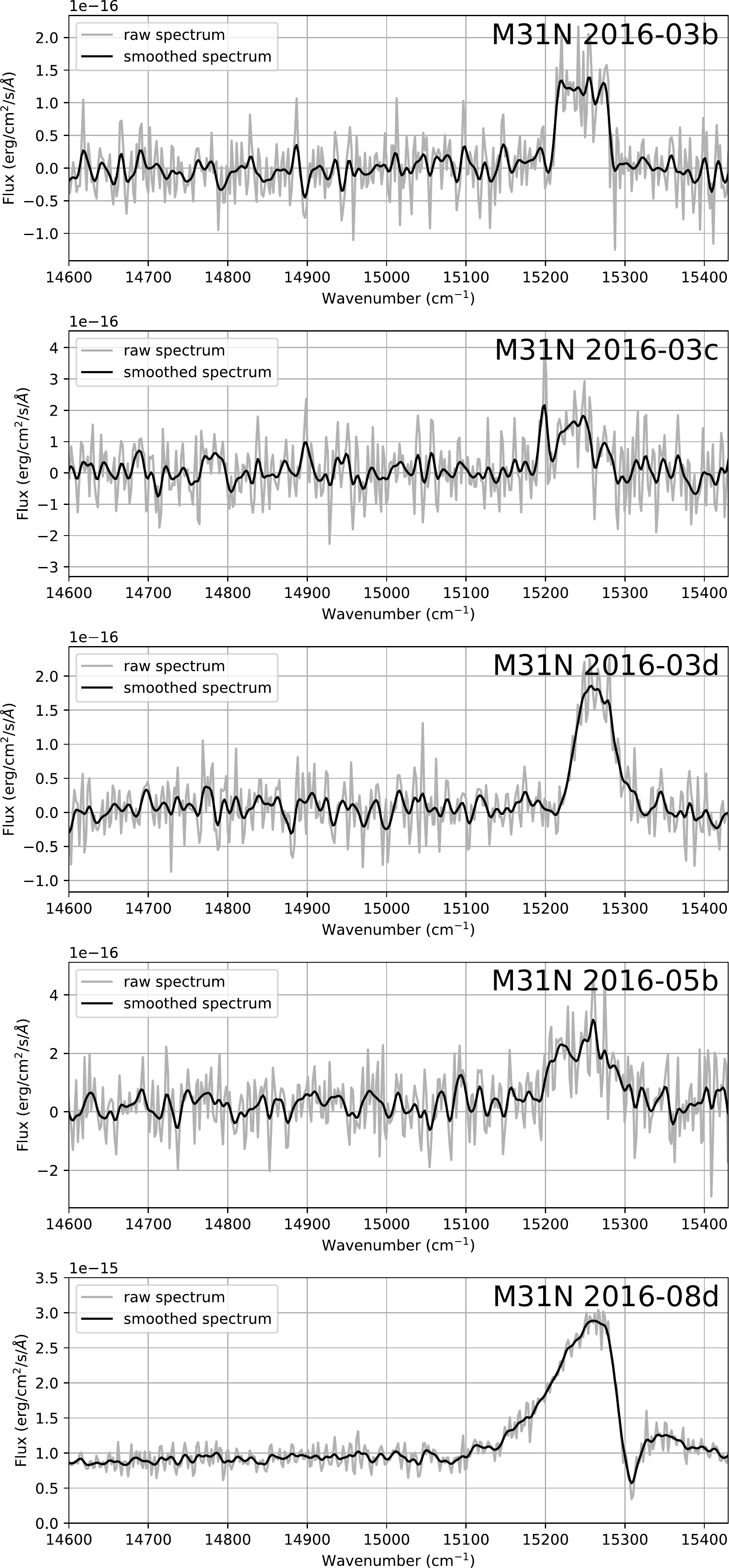}
  \caption{Novae spectra extracted from the data cube. Note that the
    integration region is smaller than the object size to enhance the
    SNR. The real flux of the nova should be considered as a lower
    limit and may be up to 50\% higher. The background-subtracted
    spectrum is plotted in grey and the fit is plotted in black.}
    \label{fig:novae}
\end{figure}

Between 65 and 100 novae erupt each year in M31
\citep{2006MNRAS.369..257D,2016MNRAS.458.2916C,2016MNRAS.455..668S},
about a third of which are detected.  Between January 11 (first nova
detected in 2016) and August 24 (the day of our observations), seven
nova candidates were detected in SITELLE's field of
view\footnote{\url{http://www.rochesterastronomy.org/sn2016/novae.html}}
but only two of them (Nova M31 2016-08d and 05b) have been
spectroscopically confirmed, both as type FeII novae.  Spectra of five
of our sources displayed an excessive broadening compared to all other
objects (see the lower right panel of Fig.~\ref{fig:lggs}): all of
them were indeed in the list of nova candidates. These objects are
listed in Table~\ref{tab:novae} but not in the catalogue because of
their transient nature; their spectra are shown in
Figure~\ref{fig:novae}.  We therefore confirm the nova nature of three
previously unconfirmed candidates (2016-03b, 03c and 03d) whereas two
candidates (2016-06b and 03e) do not display obvious nova signature in
our data.

\begin{table*}
  \caption{Novae visible in our data. The associated spectra are shown in Figure~\ref{fig:novae}.}
  \label{tab:novae}
  \begin{tabular}{lccccc}
    \hline
    ID & Detection date & Atel \#& Type & Coordinates & Reference\\
    \hline  
    M31N\,2016-03b & 2016 Feb. 26.737 UT & \#8785 & ---& 0:42:19.54\,+41:11:14.0 & \citet{Hornoch2016b}\\
    M31N\,2016-03c & 2016 Feb. 7.753 UT & \#8787 & ---& 0:42:49.25\,+41:16:40.16 & \citet{Hornoch2016c}\\
    M31N\,2016-03d & 2016 Mar. 17.768 UT & \#8838 & ---& 0:43:01.58\,+41:14:08.7 &  \citet{Hornoch2016}\\
    M31N\,2016-05b & 2016 May 28.046 UT & \#9098&Fe\,II & 0:42:42.88\,+41:15:27.1 & \citet{Darnley2016}\\
    M31N\,2016-08d & 2016 Aug. 17.102 UT & \#9388& Fe\,II & 0:42:23.53\,+41:17:19.8& \citet{Williams2016}\\
    
    \hline
  \end{tabular}
\end{table*}

\subsubsection{Supernova remnants}

Three supernova remnants (SNR) were previously known in SITELLE's
field \citep{Lee2014, Williams2004, Sasaki2012}. These SNR are all
presumably from Type Ia events since there is no signature of recent
star formation in our field of view: recent studies found no \HII{}
regions \citep{Azimlu2011} or young clusters ($<1$\,Gy)
\citep{Kang2011} and the dust seems to be only heated by the old
stellar population \citep{Viaene2014}.

\begin{figure}
  \includegraphics[width=\columnwidth]{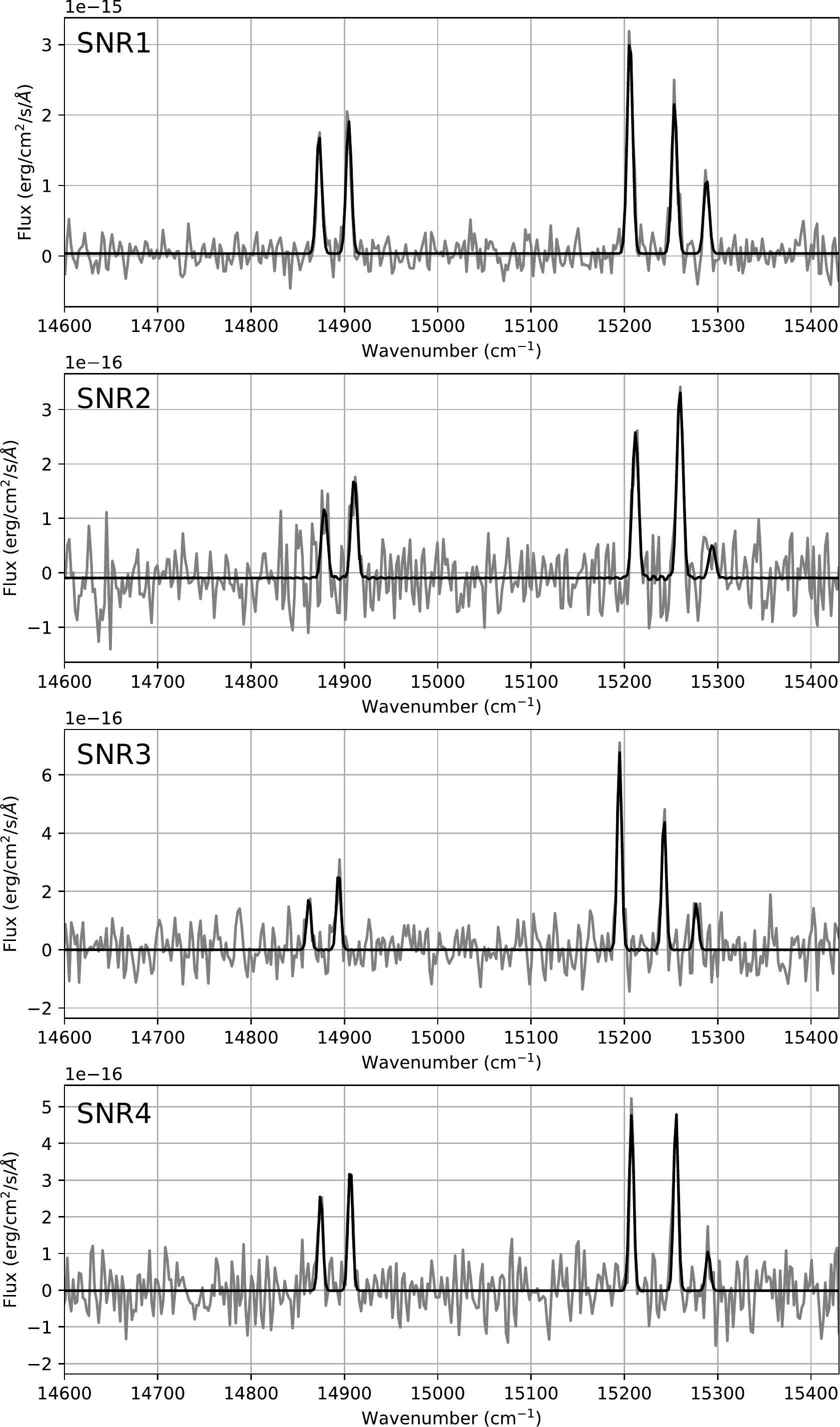}
  \caption{Integrated spectra of the supernovae remnants. The spectrum
    of the SNR3 has not been integrated over the whole visible region
    but only on its brightest knot. The background-subtracted spectrum
    is plotted in grey and the fit is plotted in black.}
    \label{fig:snr_all}
\end{figure}

They clearly stand out in our cube, and we have detected one more
(labeled SNR2). We have used the same criteria as \citet{Lee2014},
based on \citet{Braun1993}, to classify this supernova remnant: an
[SII]/H$\alpha$ ratio greater than 0.4 (0.464$\pm$0.086 in our case)
and the absence of a blue star. The basic characteristics of these
four supernova remnants are presented in table~\ref{tab:snr}. The
corresponding spectra are shown in Figure~\ref{fig:snr_all}.
\begin{table*}
  \caption{Characteristics of the observed supernova remnants.}
  \label{tab:snr}
  \begin{tabular}{lccccc}
    \hline
    ID & other ID & Coordinates & Radial velocity (km/s) & Broadening (km/s)\\
    \hline
    SNR1 & 1050$^b$ & 0:42:50.50\,+41:15:56.19& -320.5 $\pm$ 3.0&58.4$\pm$2.1\\
    SNR2 & --- & 0:42:25.07\,+41:17:34.79& -429.5 $\pm$ 4.9&59.2$\pm$4.6\\
    SNR3 & \#41$^a$, r2-57$^c$ & 0:42:24.68\,+41:17:30.79& -101.0 $\pm$ 3.8&47.2$\pm$3.1\\
    SNR4 & \#43$^a$ & 0:42:25.67\,+41:17:49.19& -351.5 $\pm$ 3.7&37.3$\pm$3.3\\
    \hline
    \multicolumn{5}{l}{$^a$ \citet{Lee2014}.}\\
    \multicolumn{5}{l}{$^b$ \citet{Stiele2011} and \citet{Sasaki2012}.}\\
    \multicolumn{5}{l}{$^c$ \citet{Williams2004}.}
  \end{tabular}
\end{table*}

\begin{figure}
  \includegraphics[width=\columnwidth]{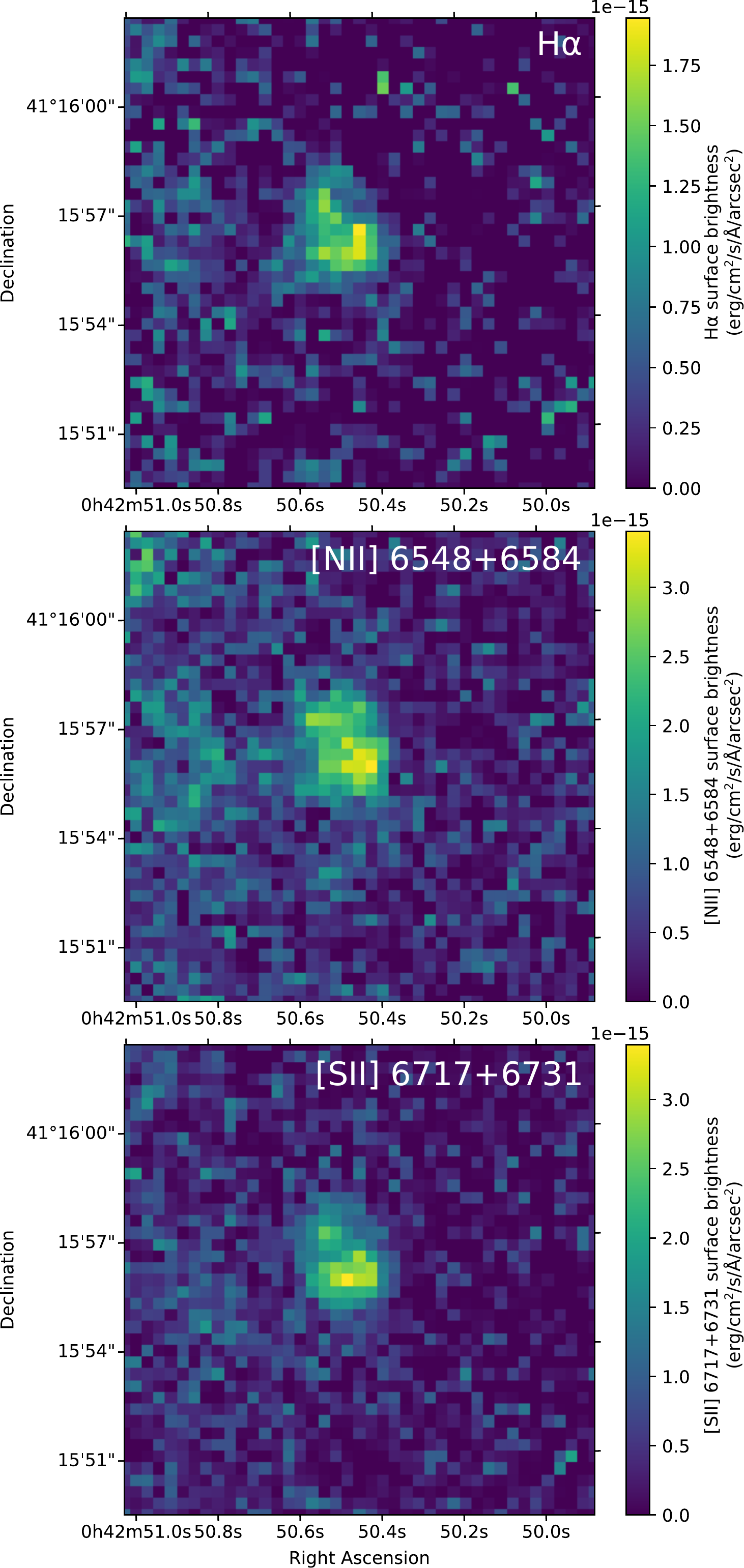}
  \caption{Flux maps in \Ha{}, \NII{} and \SII{} of the supernova
    remnant SNR1. It is catalogued as the X-ray source 1050 in
    \citet{Stiele2011}. It has been classified as a SNR and studied in
    details by \citet{Sasaki2012}.}
    \label{fig:SNR1_all}
\end{figure}

\paragraph{SNR1}
SNR1 is clearly present in the \Ha{} and \SII{} images (field 5)
provided by the Local Group Survey \citep{Massey2006} and it is
classified as a SNR by \citet{Sasaki2012} but it has not been selected
as a supernova remnant candidate by the team, although our spectrum
clearly shows strong \SII{} lines; all its lines are broader than the
instrumental resolution. A detailed analysis of this object, \#1050 in
the catalogue of \citet{Stiele2011}, in the X-ray and optical
wavelengths can be found in \citet{Sasaki2012}. They report an \Ha{}
flux of 7.5\,$10^{-15}$$\pm$4.4\,$10^{-16}$\,erg\,cm$^{-2}$\,s$^{-1}$ and a mean
\SII{} flux of 7.7\,$10^{-15}$$\pm$1.5\,$10^{-15}$\,erg\,cm$^{-2}$\,s$^{-1}$ which
compares well with our fluxes of
6.8\,$10^{-15}$$\pm$4.7\,$10^{-16}$\,erg\,cm$^{-2}$\,s$^{-1}$ for \Ha{} and
5.8\,$10^{-15}$$\pm$3.2\,$10^{-16}$\,erg\,cm$^{-2}$\,s$^{-1}$ for \SII{} measured on
the integrated spectrum shown in figure~\ref{fig:snr_all}.  Images of
its spatially resolved emission in \Ha{}, \NII{} and \SII{} are shown
in Figure~\ref{fig:SNR1_all}.

\begin{figure}
  \includegraphics[width=\columnwidth]{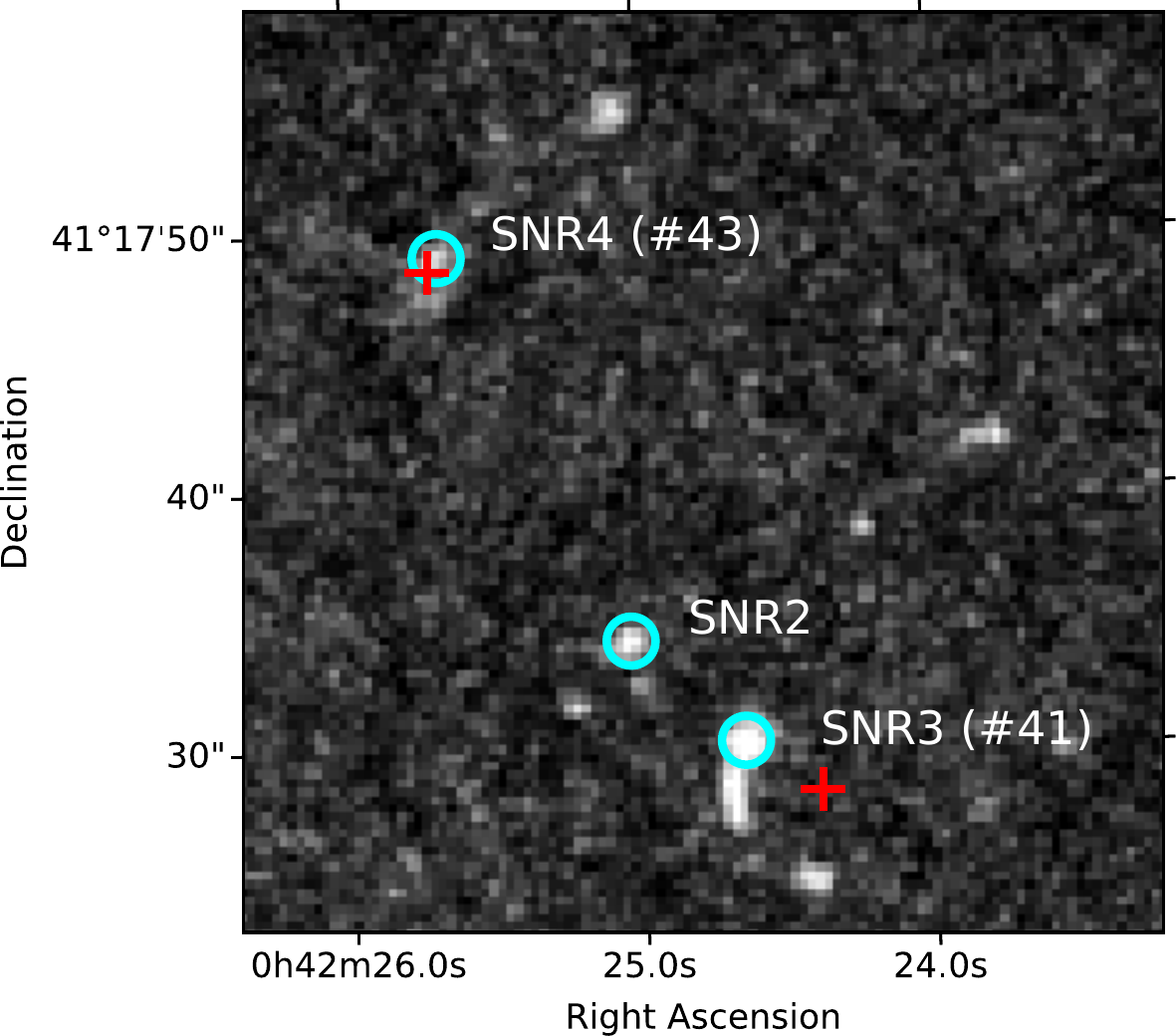}
  \caption{Small part of the detection map (see
    section~\ref{sec:detection_of_the_sources}) showing the proximity
    of 3 of the 4 supernovae remnants detected in the central region
    of M\,31. Red crosses show the position of the center of the
    remnants reported in the catalogue of \citet{Lee2014}. Blue
    circles indicate the regions integrated to compute the spectra
    shown in figure~\ref{fig:snr_all}. Note that we have chosen the
    brightest node of the remnant which has no reason to be the same
    as the central position reported by \citet{Lee2014}.}
  \label{fig:snr_region}
\end{figure}

\begin{figure}
  \includegraphics[width=\columnwidth]{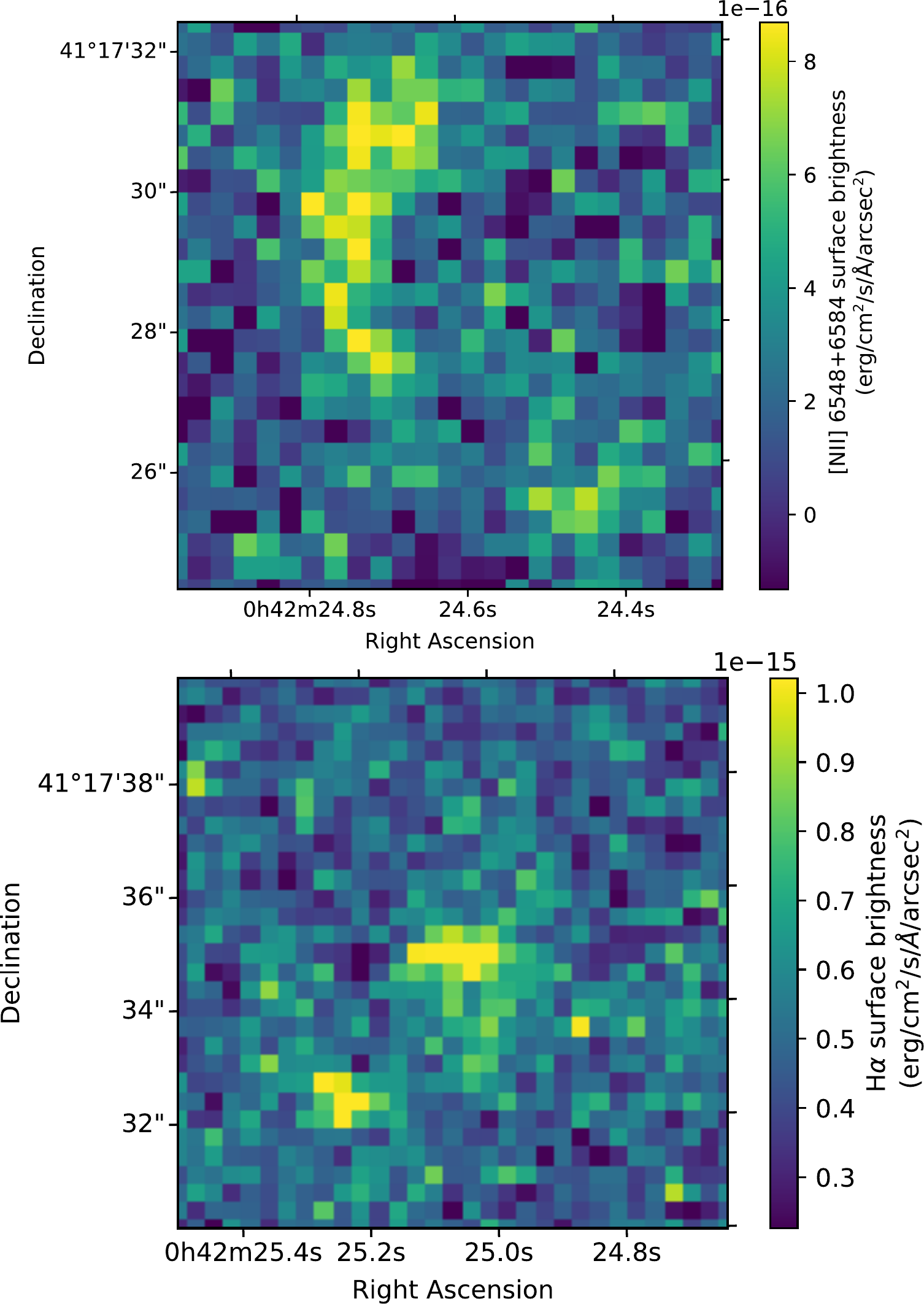}
  \caption{\textit{Top: }\NII{} surface brightness map of the
    supernova remnant SNR3 (\#41 of \citealt{Lee2014}, r2-57 in
    \citealt{Williams2004}). \textit{Bottom: } \Ha{} surface
    brightness map of the newly discovered supernova remnant SNR2.}
    \label{fig:SNR2_all}
\end{figure}

\paragraph{SNR2, SNR3 and SNR4}
Three of of the four SNRs in SITELLE's field (Lee \& Lee 41 and 43, as
well as our new SNR2) are within 30$\,\arcsec$ of each other (see
figure~\ref{fig:snr_region}). If SNR3 shows a velocity much different
from SNR2 and SNR4, the latter have a velocity difference of only
80\,\kms{} (see Table~\ref{tab:snr}), suggesting that they could be
related to the same event making the remnant 60\,pc wide which is a
little larger than the galactic supernova remnant NGC\,6960. The
detection map, which shows the highest intensity in the spectrum of
each pixel relatively to its neighbourhood is shown in
Figure~\ref{fig:snr_region}. A detailed study of the SNR3 (named r2-57
in their article) in the X-ray, optical and radio wavelengths can be
found in \citet{Williams2004}. In their Figure~3, the \SII{} emission
of the SNR2 can also be seen but nothing's found in the \OIII{} image
of the LGGS \citep{Massey2006} while the SNR3 is clearly visible.
The \Ha{} surface brightness map of SNR2 and the \NII{} surface
brightness map of SNR4 (Lee \& Lee 43), the most extended of all four
SNRs in our field ($8\,\arcsec \sim 30$ pc), are shown in
Figure~\ref{fig:SNR2_all}.

\section{Conclusions}

We have documented and demonstrated the quality of a set of general
calibration methods for SITELLE which has been applied to a spectral
cube covering the center of M\,31 in the
\Ha{}--\NII{}--\SII{} range (649--684\,nm). Although the main target
of this dataset was to study line ratios and kinematics of the diffuse
gas surrounding the core of M\,31 (Melchior et al., in preparation),
one of the first side effects of this project is the publication of
the larger and the most precise catalogue of the \Ha{}-emitting point
sources in the central region where the background flux is so high
that emission-line sources are less detectable than on the
outskirt. This is also a very interesting test since the detection of
these sources has been made in the worst possible conditions for a
Fourier transform spectrometer considering the very high continuum
background of the target. Nevertheless, the number of known
emission-line point-sources in this region is more than two times
larger than the previously published catalogues, 5 novae spectra have
been obtained as long as spatially resolved data of 4 supernovae
remnants, one of them being a new discovery. Great care was taken to
compare the obtained data with the best catalogues
\citep{Merrett2006,Halliday2006} and images \citep{Massey2006} of this
region which demonstrate the quality of the calibration as long as the
completeness of the catalogue. A deeper analysis of this catalogue
combined with the data obtained in the \OIII--H$\beta$ range will be
made in the subsequent articles. We believe that the quality of the
calibration made will prove useful for all the other studies made with
this data cube.

\section*{Acknowledgements}

Based on observations obtained with SITELLE, a joint project of
Universit\'e Laval, ABB, Universit\'e de Montr\'eal and the
Canada-France-Hawaii Telescope (CFHT) which is operated by the
National Research Council (NRC) of Canada, the Institut National des
Science de l'Univers of the Centre National de la Recherche
Scientifique (CNRS) of France, and the University of Hawaii. LD is
grateful to the Natural Sciences and Engineering Research Council of
Canada, the Fonds de Recherche du Qu\'ebec, and the Canada
Foundation for Innovation for funding.  This research has made use
of Astropy, a community-developed core Python package for Astronomy
(Astropy Collaboration, 2013, \url{http://www.astropy.org}). This
work has made use of data from the European Space Agency (ESA)
mission {\it Gaia} , processed by the {\it Gaia} Data Processing and
Analysis Consortium (DPAC). Funding for the DPAC has been provided
by national institutions, in particular the institutions
participating in the {\it Gaia} Multilateral Agreement.  Special
thanks to Simon Prunet for his suggestions regarding the improvement
of the robustness of the first-level registration algorithm. We are
also thankful to Simon Thibault and Denis Brousseau for their
helpful comments on the modeling of the calibration laser map, as well as to the anonymous referee who provided 
a detailed feedback and constructive criticism which helped improve the quality of this paper.





\bibliographystyle{mnras}
\bibliography{m31sn3} 




\appendix

\section{Relation between the incident angle and the measured wavenumber}

\begin{figure}
  \includegraphics[width=\columnwidth]{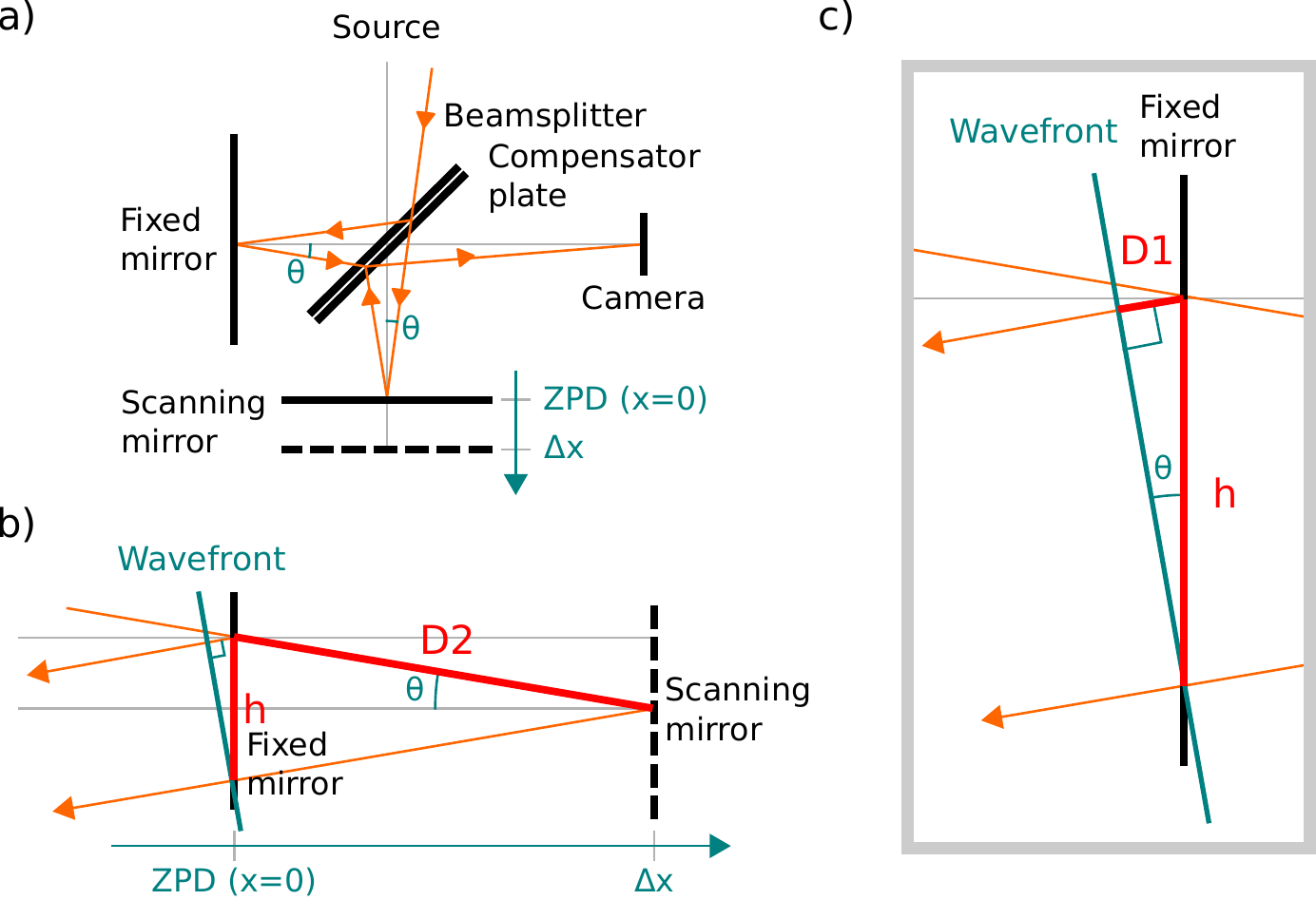}
  \caption{\textit{a)} Optical path of the light at an incident angle
    $\theta$ in a classical Michelson interferometer. \textit{b)}
    Superposition of the optical path followed by the light in the
    fixed mirror arm and the moving mirror arm. The moving mirror is
    at a mechanical distance $\Delta x$ from the ZPD (the position
    where both mirrors are at the same distance of the
    beamsplitter). $D2$ is the optical distance crossed by light along
    the moving mirror path. \textit{c)} Zoomed-in portion of figure b)
    showing the optical distance $D1$ crossed by light along the fixed
    mirror path.}
    \label{fig:incident_angle}
\end{figure}

When light enters a Michelson interferometer at a null incident angle
(see Figure~\ref{fig:incident_angle}), the optical path difference
(OPD) $x$ is simply two times the mechanical distance $\Delta x$
between the actual position of the moving mirror and the position
where both mirrors are at the same distance of the beamsplitter (known
as the zero path difference position, ZPD):
\begin{equation}
  x = 2 \Delta x
\end{equation}
When the incident angle $\theta$ is different from 0 the OPD is known
to be (see e.g. \citealt{Davis2001} p.~70)
\begin{equation}
  \label{eq:opd_cos_theta}
  OPD = x \cos(\theta)
\end{equation}
For the sake of clarity and for future reference we reproduce here the
demonstration of this relation as given in the thesis of
\citet{Grandmont2006}.

The OPD at any given incident angle is, by definition, the difference
between the distance crossed by light in the fixed mirror arm ($D1$)
and the moving mirror arm ($2\times D2$) measured on the wavefront,
i.e. perpendicularly to the light direction:
\begin{equation}
  OPD = 2\times D2 - D1\;.
\end{equation}
A superposition of both optical paths is shown on
Figure~\ref{fig:incident_angle}. From simple geometrical consideration
we can deduce that $D2=\Delta x / \cos(\theta)$,
$h=2\,D2\,\sin(\theta)$ and $D1=h \sin(\theta)$ from which it follows
that
\begin{equation}
  D1 = 2\,D2\,\sin(\theta)^2\;.
\end{equation}
We can finally write
\begin{equation}
  OPD = 2 \frac{\Delta x}{\cos(\theta)}\left(1 - \sin(\theta)^2\right) = 2\Delta x \cos(\theta) = x\cos(\theta)
\end{equation}

\begin{figure}
  \includegraphics[width=\columnwidth]{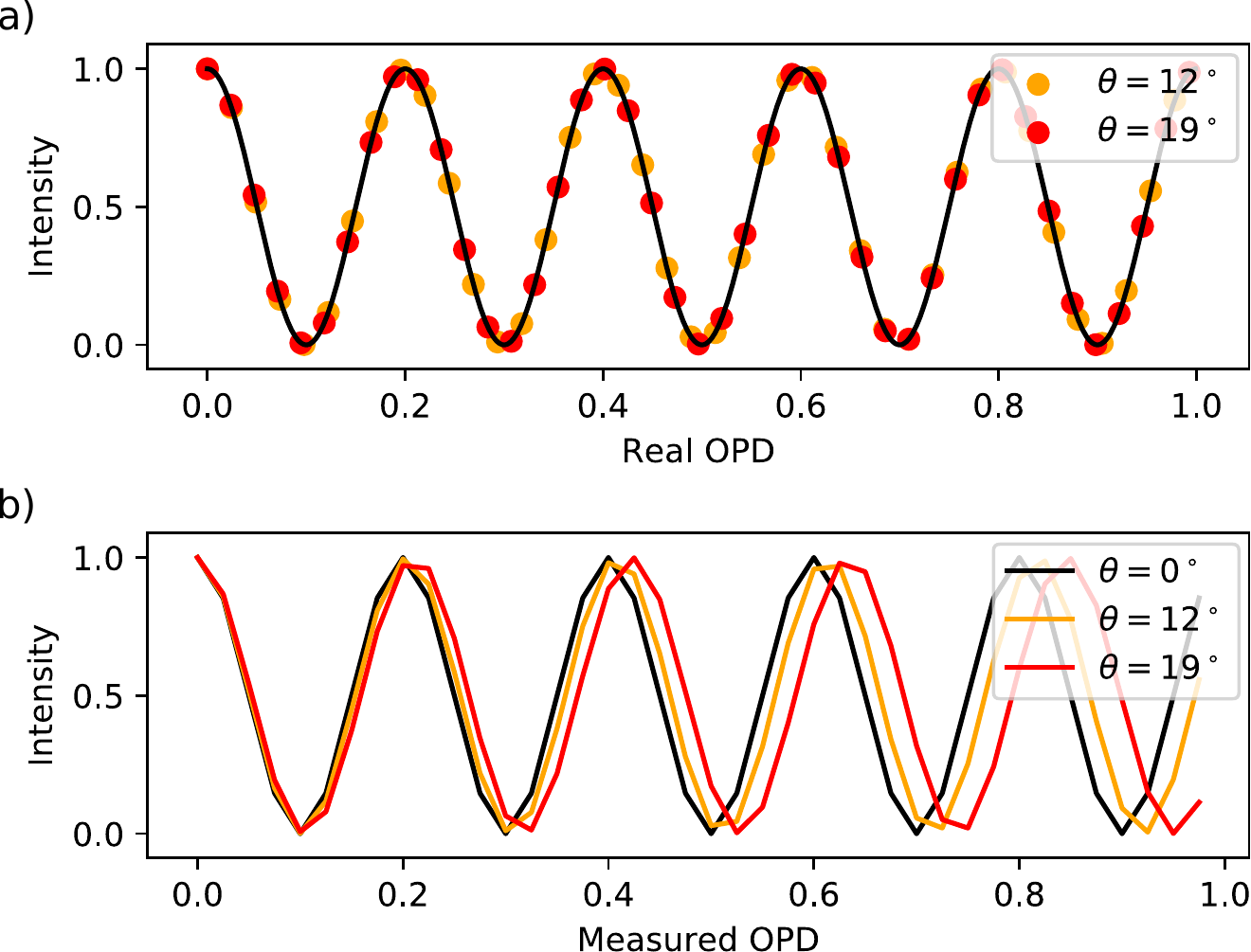}
  \caption{Effect of the incident angle $\theta$ on the sampling
    step-size and the measured frequency of an
    emission-line. \textit{a)} Positions of the recorded samples of
    the interferogram of a monochromatic line at two incident angles
    12$^{\circ}$ and 19$^{\circ}$ (corresponding roughly to the top
    and bottom pixels of the field of view). \textit{b)} Recorded
    interferograms at three different incident angles (on-axis and at
    12$^{\circ}$ and 19$^{\circ}$). The measured OPD of each sample,
    which differs from the real OPD, is the on-axis OPD i.e. the
    target OPD of the mirror controller. The variation of the measured
    wavelength of the emission line appears clearly.}
    \label{fig:thetaex_mod}
\end{figure}

The effect on the measured wavenumber $\sigma$ of a single emission-line of real wavenumber $\sigma_0$ is
illustrated on Figure~\ref{fig:thetaex_mod}. As the OPD of the light
coming with an incident angle is smaller, the interferogram of the
source is sampled at shorter steps. If the centroid of the line is
calculated considering the step size measured on-axis by the mirror
controller the same emission-line will thus see its observed
wavenumber increased by a factor $1/\cos(\theta)$ i.e.:
\begin{equation}
  \label{eq:sigma_cos_theta}
  \sigma = \frac{\sigma_0}{\cos\theta}
\end{equation}

\section{Excerpt from the catalogue}

\begin{landscape}
  \begin{table}
    \tiny
    \caption{Excerpt from the catalogue.}
    \label{tab:catalog_example}
    \begin{tabular}{lccccccccccccccccccccc}
      ID&$\alpha_{2000}$&$\delta_{2000}$&$V$$^a$&$\epsilon_V$$^a$&$\sigma$$^a$&$\epsilon_{\sigma}$$^a$&$F$(\Ha{})$^b$&$\epsilon_{F(\text{H}\alpha)}$$^b$&$F(\text{\NII{}}\,\lambda 6584$)$^b$&$\epsilon_{F(\text{\NII{}}\,\lambda 6584)}$$^b$&$F(\text{\SII{}}$)$^b$&$\epsilon_{F(\text{\SII{}})}$$^b$&Me06$^c$&Ha06$^d$&Ma06 \Ha{}$^e$&Ma06 \OIII$^e$&Comments\\
      \hline
S004236.99+411209.3&00 42 36.99&+41 12 09.3&-0595.6&+0002.4&+0032.9&+0001.0&2.35e-15&1.23e-16&1.78e-15&1.00e-16&<4.1e-16&---&2806&P69       &Sat.&Sat.&---\\
S004239.81+411641.5&00 42 39.81&+41 16 41.5&-0175.5&+0002.5&+0020.0&+0001.9&2.30e-15&1.64e-16&1.73e-15&1.41e-16&<6.3e-16&---&---&---&Sat.&Sat.&---\\
S004241.24+411909.5&00 42 41.24&+41 19 09.5&-0144.7&+0002.7&+0011.6&+0004.1&8.51e-16&7.13e-17&<1.2e-16&---&<2.1e-16&---&802&P105      &Sat.&Sat.&---\\
S004254.20+411336.8&00 42 54.20&+41 13 36.8&-0341.0&+0002.5&+0020.6&+0001.9&1.24e-15&8.20e-17&<2.2e-16&---&<1.5e-16&---&1241&P475      &Sat.&Sat.&---\\
S004242.16+411820.3&00 42 42.16&+41 18 20.3&+0045.4&+0004.2&+0057.4&+0003.6&2.25e-15&2.08e-16&<5.1e-16&---&<2.6e-16&---&---&---&Sat.&Sat.&---\\
S004229.73+411351.8&00 42 29.73&+41 13 51.8&-0504.5&+0002.3&+0022.4&+0000.8&4.26e-15&1.96e-16&1.55e-15&9.51e-17&<5.0e-16&---&1131&P98       &Sat.&Sat.&---\\
S004254.94+411257.2&00 42 54.94&+41 12 57.2&-0438.5&+0002.3&+0023.1&+0001.0&1.80e-15&9.47e-17&1.48e-15&8.20e-17&<3.2e-16&---&---&---&Sat.&Sat.&---\\
S004244.51+411813.4&00 42 44.51&+41 18 13.4&-0022.9&+0002.5&+0017.1&+0002.1&1.86e-15&1.24e-16&7.56e-16&8.11e-17&<3.4e-16&---&3029&---&Sat.&Sat.&---\\
S004302.60+411558.3&00 43 02.60&+41 15 58.3&-0040.2&+0002.3&+0019.4&+0001.2&1.74e-15&9.89e-17&1.79e-15&1.01e-16&<4.1e-16&---&---&---&Sat.&Sat.&---\\
S004235.97+411410.5&00 42 35.97&+41 14 10.5&-0280.7&+0002.6&+0019.9&+0002.2&1.29e-15&1.04e-16&9.87e-16&9.10e-17&<2.7e-16&---&1205&---&Sat.&Sat.&---\\
S004257.87+411243.4&00 42 57.87&+41 12 43.4&-0186.2&+0002.3&+0028.1&+0000.9&2.52e-15&1.23e-16&4.86e-16&5.09e-17&<2.3e-16&---&1232&---&Sat.&Sat.&---\\
S004257.70+411951.0&00 42 57.70&+41 19 51.0&-0324.2&+0002.3&+0016.5&+0001.4&1.69e-15&9.28e-17&1.00e-15&6.69e-17&<2.8e-16&---&809&---&Sat.&Sat.&---\\
S004239.73+411839.4&00 42 39.73&+41 18 39.4&-0362.3&+0002.4&+0021.1&+0001.4&1.92e-15&1.10e-16&3.83e-16&5.66e-17&<3.2e-16&---&---&---&Sat.&Sat.&---\\
S004255.48+411721.1&00 42 55.48&+41 17 21.1&-0287.6&+0002.4&+0034.2&+0000.9&3.79e-15&1.97e-16&2.94e-15&1.63e-16&7.61e-16&1.33e-16&1268&P30       &Sat.&Sat.&---\\
S004245.85+411735.4&00 42 45.85&+41 17 35.4&+0020.3&+0002.6&+0022.9&+0002.0&1.75e-15&1.36e-16&1.37e-15&1.20e-16&<6.4e-16&---&1360&---&Sat.&Sat.&---\\
S004242.48+411356.4&00 42 42.48&+41 13 56.4&-0413.6&+0002.7&+0029.9&+0001.7&1.90e-15&1.36e-16&1.20e-15&1.05e-16&<5.7e-16&---&1246&---&Sat.&Sat.&---\\
S004231.48+411614.7&00 42 31.48&+41 16 14.7&-0378.7&+0002.6&+0020.1&+0002.2&1.44e-15&1.07e-16&<3.1e-16&---&<2.3e-16&---&1162&---&Sat.&Sat.&---\\
S004239.57+411831.6&00 42 39.57&+41 18 31.6&-0261.0&+0002.7&+0021.7&+0002.4&1.23e-15&9.74e-17&<2.9e-16&---&<2.4e-16&---&3270&P102      &Sat.&Sat.&---\\
S004235.29+411446.0&00 42 35.29&+41 14 46.0&-0255.4&+0002.3&+0022.3&+0000.9&4.59e-15&2.21e-16&7.73e-16&8.88e-17&<3.2e-16&---&1139&P53       &Sat.&Sat.&---\\
S004246.10+411516.5&00 42 46.10&+41 15 16.5&-0192.7&+0002.9&+0028.3&+0002.1&2.43e-15&1.92e-16&<4.8e-16&---&<1.9e-16&---&1287&---&Sat.&Sat.&---\\
S004256.13+411632.5&00 42 56.13&+41 16 32.5&-0172.2&+0002.5&+0020.0&+0001.7&1.57e-15&1.10e-16&1.53e-15&1.09e-16&<3.7e-16&---&---&---&Sat.&Sat.&---\\
S004305.91+411846.0&00 43 05.91&+41 18 46.0&-0550.1&+0002.5&+0020.1&+0001.7&1.38e-15&8.84e-17&7.63e-16&6.43e-17&<2.8e-16&---&---&---&Sat.&Sat.&---\\
S004232.73+411916.9&00 42 32.73&+41 19 16.9&-0518.7&+0002.3&+0019.3&+0000.9&2.67e-15&1.25e-16&<2.2e-16&---&<2.0e-16&---&2753&---&Sat.&Sat.&---\\
S004229.58+411404.3&00 42 29.58&+41 14 04.3&-0275.5&+0002.4&+0022.0&+0001.2&2.34e-15&1.26e-16&9.41e-16&7.36e-17&<3.3e-16&---&1132&P50       &Sat.&Sat.&---\\
S004251.86+411214.7&00 42 51.86&+41 12 14.7&-0555.6&+0002.3&+0025.3&+0000.7&3.53e-15&1.58e-16&<2.3e-16&---&<3.4e-16&---&1225&---&Sat.&Sat.&---\\
      S004302.58+412028.3&00 43 02.58&+41 20 28.3&-0257.9&+0002.4&+0020.5&+0001.5&1.35e-15&7.99e-17&4.76e-16&4.77e-17&<1.9e-16&---&2755&P200      &Sat.&Sat.&elongated PSF\\
      \hline
      \multicolumn{18}{l}{$^a$ Velocity $V$ and broadening $\sigma$ are expressed in \kms{}. The uncertainties are respectively $\epsilon_V$ and $\epsilon_\sigma$.}\\
      \multicolumn{18}{l}{$^b$ Fluxes are expressed in erg\,cm$^{-2}$\,s$^{-1}$. The uncertainties are represented with the letter $\epsilon$. In the case of \SII{}, the reported flux is the sum of \SII{}\,$\lambda 6717$ and \SII{}\,$\lambda 6731$; in most cases only an upper limit can be given.} \\
      \multicolumn{18}{l}{$^c$ Index of the cross-matched source of \citet{Merrett2006}.}\\
      \multicolumn{18}{l}{$^d$ Index of the cross-matched source of \citet{Halliday2006}.}\\
      \multicolumn{18}{l}{$^e$ Index of the cross-matched source of \citet{Massey2006} in the \OIII{}/B frame or the \Ha{}/R frame (see text for details). (T) indicates a detection ; (Sat.) indicates a saturated region.}

    \end{tabular}
  \end{table}
\end{landscape}



\bsp	
\label{lastpage}
\end{document}